\documentclass[structabstract]{aa}  
\usepackage{graphicx}
\usepackage{txfonts}
\usepackage{natbib}
\bibpunct{(}{)}{;}{a}{}{,}
\usepackage{amssymb}
\usepackage{xcolor}
\usepackage{lipsum}
\usepackage{textcomp}
\usepackage{multirow}
\usepackage[colorlinks=true,linkcolor=blue,citecolor=blue,
bookmarks=true,bookmarksopen=true,bookmarksnumbered=true,draft=false]{hyperref}
\usepackage{enumerate}

\newcommand{\msun}{M$_{\sun}$}
\newcommand{\xmm}{{\it XMM-Newton}}

\newcommand{\hour}{$^{\mathrm{h}}$}
\newcommand{\minute}{$^{\mathrm{m}}$}
\newcommand{\second}{$^{\mathrm{s}}$}
\newcommand{\snra}{MCSNR J0508$-$6830}
\newcommand{\snrb}{MCSNR J0511$-$6759}
\newcommand{\snrc}{MCSNR J0514$-$6840}
\newcommand{\snrd}{MCSNR J0517$-$6759}

\newcommand{\siiha}{[\ion{S}{ii}]/H$\alpha$}

\begin{document}

   \title{Four new X-ray-selected supernova remnants\\ in the Large
Magellanic Cloud\,\thanks{Based on observations
obtained with \xmm, an ESA science mission with instruments and contributions
directly funded by ESA Member States and NASA.}}

   \subtitle{}

    \titlerunning{Four new X-ray-selected supernova remnants in the LMC}

    \author{P. Maggi    \inst{1}
    \and F.    Haberl   \inst{1}
    \and P. J. Kavanagh \inst{2}
    \and S. D. Points \inst{3}
    \and J.    Dickel \inst{4}
    \and L. M. Bozzetto \inst{5}
    \and\\ M.    Sasaki \inst{2}
    \and Y.-H. Chu \inst{6}
    \and R. A. Gruendl \inst{6}
    \and M. D. Filipovi\'c \inst{5}
    \and W.    Pietsch \inst{1}
    }

   \institute{Max-Planck-Institut f\"ur extraterrestrische Physik, Postfach
    1312, Giessenbachstr., 85741 Garching, Germany\\ \email{pmaggi@mpe.mpg.de}
    \and
    Institut f\"ur Astronomie und Astrophysik T\"ubingen, Universit\"at
    T\"ubingen, Sand 1, 72076 T\"ubingen, Germany
    \and
    Cerro Tololo Inter-American Observatory, National Optical
    Astronomy Observatory, Cassilla 603 La Serena, Chile 
    \and
    Physics and Astronomy Department, University of New Mexico, MSC 07-4220,
    Albuquerque, NM 87131, USA
    \and
    University of Western Sydney, Locked Bag 1797, Penrith South DC, NSW 1797,
    Australia
    \and
    Astronomy Department, University of Illinois, 1002 West Green Street,
    Urbana, IL 61801, USA
    }

   \date{Received \,/\,Accepted }

  \abstract
    {}
    {We present a detailed multi-wavelength study of four new supernova
remnants (SNRs) in the Large Magellanic Cloud (LMC). The objects were
identified as SNR candidates in X-ray observations performed during the survey
of the LMC with \xmm.
}
    {Data obained with \xmm\ are used to investigate the morphological and
spectral features of the remnants in X-rays. We measure the plasma conditions,
look for supernova (SN) ejecta emission, and constrain some of the SNR
properties (e.g. age and ambient density). We supplement the X-ray data with
optical, infrared, and radio-continuum archival observations, which allow us to
understand the conditions resulting in the current appearance of the remnants.
Based on the spatially-resolved star formation history (SFH) of the LMC
together with the X-ray spectra, we attempt to type the supernovae that created
the remnants.
}
    {We confirm all four objects as SNRs, to which we assign the names \snra,
\snrb, \snrc, and \snrd. In the first two remnants, an X-ray bright plasma is
surrounded by very faint [\ion{S}{ii}] emission. The emission from the central
plasma is dominated by Fe L-shell lines, and the derived iron abundance is
greatly in excess of solar. This establishes their type Ia (i.e. thermonuclear)
SN origin. They appear to be more evolved versions of other Magellanic Cloud
iron-rich SNRs which are centrally-peaked in X-rays. From the two other remnants
(\snrc\ and \snrd), we do not see ejecta emission. At all wavelengths at which
they are detected, the local environment plays a key role in their observational
appearance. We present evidence that \snrd\ is close to and interacting with a
molecular cloud, suggesting a massive progenitor.
}
    {}


   \maketitle


\begin{table*}[ht]
\caption{\xmm\ observations log, position, and size of the remnants.}
\begin{center}
\label{table_xray_observations}
\begin{tabular}{l c c c c c c c}
\hline
\hline
\noalign{\smallskip}
MCSNR & Observation & \multicolumn{2}{c}{Exposure time (ks)\tablefootmark{a}}
&
\multicolumn{2}{c}{Position of the remnant (J2000)} & X-ray
radius\tablefootmark{b} & Off-axis \\
 & ID & pn & MOS1/2 & RA & DEC & (arcmin) &
angle\tablefootmark{c}\\
\noalign{\smallskip}
\hline
\noalign{\smallskip}
A: J0508$-$6830 &  0690742401 & 26.0 & 27.2 & 05\hour\,08\minute\,49.5\second
&
$-$68\degr\,30\minute\,41\second & $1.15 (\pm 0.10)\times 0.90 (\pm 0.10)$ &
10.9 \\
\noalign{\smallskip}
B: J0511$-$6759 & 0690742201 & 25.1 & 28.9 & 05\hour\,11\minute\,10.7\second &
$-$67\degr\,59\minute\,07\second & $0.93 \pm 0.09$ & 7.9 \\
\noalign{\smallskip}
\noalign{\smallskip}
\multirow{2}{*}{C: J0514$-$6840} & 0690742601 & 27.3 & 28.0 & \multirow{2}{*}{
05\hour\,14\minute\,15.5\second } & \multirow{2}{*}{
$-$68\degr\,40\minute\,14\second} & \multirow{2}{*}{ $1.83\pm0.12$ } & 10.0 \\
 & 0690742701 & 29.5 & 33.6 & & & & 13.4 \\
\noalign{\smallskip}
\noalign{\smallskip}
D: J0517$-$6759 & 0690741101 & 24.6 & 26.2 & 05\hour\,17\minute\,10.2\second &
$-$67\degr\,59\minute\,03\second & $2.7 (\pm 0.1) \times 1.8 (\pm 0.1)$ & 4.1 \\
\noalign{\smallskip}
\hline
\end{tabular}
\end{center}
\tablefoottext{a}{Exposure times after removal of high background intervals.}
\tablefoottext{b}{See Sect.\,\ref{results_morphology} for details on the size of
each remnant, or size at other wavelengths.}
\tablefoottext{c}{Angle in arcmin between the aiming point of the observations
and the centre of the X-ray source (as defined in
Sect.\,\ref{results_morphology}).}
\end{table*}

\section{Introduction}
\label{introduction}

The study of supernova remnants (SNRs) is crucial to advance our understanding
of many important astrophysical processes. SNRs mark the end point of stellar
evolution, and return nucleosynthesis products to the interstellar medium (ISM),
enriching it with heavy elements. The tremendous amount of energy (\mbox{$\sim
10^{51}$} erg) of supernova (SN) explosions heats the ISM up to X-ray emitting
temperature ($> 10^6$ K) and contributes to the mixing of the freshly-produced
elements in the SNR environment. Cosmic rays, another important part of the ISM
energy budget, can be accelerated up to $10^{18}$ eV in SNR shocks
\citep{2001MNRAS.321..433B}.

Many questions related to SNe remain open. The exact nature of the progenitor of
type Ia, i.e. thermonuclear SNe, being either a white dwarf accreting from a
companion or a merger of two white dwarves, is hotly debated
\citep[e.g.][and references therein]{2008MNRAS.384..267M}. Core-collapse (CC)
SNe mark the death of massive, short-lived stars. The lower limit on the mass of
a star undergoing a CC SN is about 8~M$_{\sun}$, but the exact value remain
elusive, and so does the mass range leading to various CC SN subtypes
\citep[see][for a review]{2009ARA&A..47...63S}. From a statistically significant
sample of SN \emph{remnants}, we can aim to obtain clues to some of these
outstanding questions.

The two flavours of SNe deposit a similar amount of energy in the ISM, producing
remnants which are harder to type the older they are. The most secure typing
methods are the study of SN light echoes
\citep{2005Natur.438.1132R,2008ApJ...680.1137R}, the measurement of the
nucleosynthesis products in the ejecta \citep[e.g. ][]{1995ApJ...444L..81H}, or
the association with a neutron star/pulsar wind nebula. These methods work well
for relatively young remnants ($\lesssim 5000$ yr), leaving a significant
fraction of the SNR population untyped. However, several \emph{evolved} SNRs
have been discovered (in X-rays) in the Magellanic Clouds with an iron-rich,
centrally bright emission
\citep{2001PASJ...53...99N,2003ApJ...593..370H,2004A&A...421.1031V,
2006ApJ...640..327S,2006ApJ...652.1259B,bozzetto2013}, naturally leading to
their classification as type Ia remnants. In addition, studies of the X-ray and
infrared morphologies of SNRs \citep{2009ApJ...706L.106L,2013ApJ...771L..38P}
suggest that, as a class, type Ia and CC SNRs have distinct symmetries: type Ia
remnants are more spherical and mirror symmetric than the CC SNRs.

Finally, clues to the type of remnants are provided by the study of the stellar
population around the SNRs. High-mass stars (i.e. CC-SN progenitors) are rarely
formed in isolation but cluster in OB associations. \citet{1988AJ.....96.1874C}
used this method to tentatively type all Large Magellanic Cloud (LMC) remnants
known at the time. With the availability of the star formation history (SFH) map
of the LMC \citep{2009AJ....138.1243H},
derived from resolved stellar populations, it is now possible to study the
connection between remnants and the age of their parent populations.
\citet{2009ApJ...700..727B} performed such a study on (young) SNRs having secure
type Ia or CC classifications. As expected, given the short lifetimes of massive
progenitors, they found that all CC SNRs in their sample had SFHs dominated by
recent peaks of star formation rate. This appears to be \emph{necessary}, but
not \emph{sufficient}. Indeed, type Ia SNRs can also be found in star-forming
regions, as they showed for N103B \citep[see
also][]{1995ApJ...444L..81H,2003ApJ...582..770L} or for SNR 0104$-$72.3 in the
SMC \citep{2011ApJ...731L...8L}. On the other hand, the type Ia SNRs of the
(limited) sample of \citet{2009ApJ...700..727B} are associated with a variety
of environments, and future similar studies will provide more insights. We
investigate the local SFHs of the new SNRs presented in this work as an
additional tool to study the origin of the remnants.

Studies of SNRs in the Milky Way are plagued by the large distance uncertainties
of, and interstellar absorption towards, sources in the Galactic plane. On the
other hand, the LMC offers an ideal laboratory for such studies: First, the
distance towards the LMC, our closest star-forming neighbour, is relatively
small and very well studied \citep[e.g.][and references
therein]{2013Natur.495...76P}. Second, the moderate inclination angle and small
line-of-sight depth of the LMC
\citep[e.g.][]{2012AJ....144..106H,2013A&A...552A.144S} mean that we can assume
all LMC sources to be at a very similar distance. Third, the interstellar
absorption by gas in the foreground is much smaller towards the LMC than towards
the Galactic plane, allowing observations even in the soft X-ray regime, below 1
keV. Finally, a wealth of data is available for the LMC from radio to X-rays,
allowing for easier detection and classification of SNRs, which classically
exhibit three signatures\,: thermal X-ray emission in the (0.2--5~keV) band,
optical line emission with enhanced [\ion{S}{ii}] to H$\alpha$ ratios, and
non-thermal (synchrotron) radio-continuum emission. For all these reasons, we
can attempt to discover and study the \emph{complete} sample of SNRs in the LMC.

The LMC is particularly rich in SNRs. \citet{2010MNRAS.407.1301B} compiled a
catalogue of 54 SNRs in the LMC. The bulk of the population was
discovered/confirmed by early X-ray and radio surveys
\citep{1981ApJ...248..925L,1983ApJS...51..345M}, and then regularly
increased with the completion of more recent surveys
(e.g. with Parkes, \citealt{1998A&AS..130..421F}; \emph{ROSAT},
\citealt{1999ApJS..123..467W}; MCELS, \citealt{2005AAS...207.2507S}; and ATCA
\citealt{2008MNRAS.383.1175P}). Although a significant fraction of the SNR
population of the LMC should have been discovered, there might still be some
fainter and more evolved remnants to be found, and candidates to
be confirmed. In particular, SNRs lacking one or two of the conventional
signatures are harder to identify. Examples of such SNRs can be found in
\citet{2010ApJ...725.2281K}, \citet{2010A&A...518A..35C},
\citet{2012A&A...539A..15G}, and \citet{2013MNRAS.432.2177B}.

We carried out a survey of the inner $\sim$4.5\degr$\times$4.5\degr\ of the LMC
with \xmm\ (Haberl et al., in prep.). In addition to unraveling the X-ray point
source population of the LMC (down to $\sim 2 \times 10^{33}$ erg\,s$^{-1}$) and
studying its hot gas content, one of the main goals of the survey is to discover
\emph{new} SNRs and to study those, already confirmed, that have not been
observed with a modern X-ray observatory. First results from the survey related
to that theme have been presented in \citet{2012A&A...546A.109M} and
\citet{bozzetto2013}.

In this paper we report the discovery of four new SNRs in the \xmm\ survey
of the LMC. A description of the survey, and of the optical, infrared, and radio
observations that we used to study the remnants is given in
Sect.\,\ref{observations}. The reasons why we selected these sources are
presented in Sect.\,\ref{sample_selection}, then the data analysis is described
in Sect.\,\ref{data}. Results are presented and discussed in
Sects.\,\ref{results} and \ref{discussion}. We summarise the results in
Sect.\,\ref{summary}. Unless otherwise stated, the errors quoted are at the
90\,\% confidence level. Throughout the paper we assume a distance to the LMC of
50 kpc \citep{2013Natur.495...76P}.

\section{Observations}
\label{observations}

\subsection{X-ray}
\label{observations_xray}
The sources presented in this work were discovered in observations performed
during the \xmm\ survey of the LMC (Haberl et al., in prep.; see
Sect.\,\ref{sample_selection}). For this survey we use the European Photon
Imaging Camera (EPIC), comprising a pn CCD imaging camera
\citep{2001A&A...365L..18S} and two MOS CCD imaging cameras
\citep{2001A&A...365L..27T}, as the main instrument. All observations are
performed using the same instrumental setting: all cameras are operated in
full-frame mode, with the thin and medium optical filter for pn and MOS cameras,
respectively.

The data used in this paper have been processed with the \xmm\
SAS\,\footnote{Science Analysis Software, \url{http://xmm.esac.esa.int/sas/}}
version 11.0.1. To screen out periods of high background activity, we applied a
threshold of 8 and 2.5 cts\,ks$^{-1}$\,arcmin$^{-2}$ on pn and MOS light curves
in the 7--15 keV energy band. This yielded various useful exposure times,
between $\sim 22$ and $\sim 33$ ks. Details of the \xmm\ observations are
listed in Table\,\ref{table_xray_observations}.

X-ray source detection is performed on all observations from our \xmm\ survey
(which will be presented in detail in future works), using the SAS task
\texttt{edetectchain} applied simultaneously to images in five energy bands
\citep[given in][Table~3]{2009A&A...493..339W}. The detection lists were
primarily used to identify our sample of four SNR candidates
(Sect.\,\ref{sample_selection}). In addition, we could look for central compact
objects (CCOs) and exclude unrelated point sources from the spectrum extraction
regions.

\subsection{Optical}
\label{observations_optical}
We made use of data from the Magellanic Clouds Emission Line Survey
\citep[MCELS, e.g.][]{2000ASPC..221...83S}. Observations were taken with the
0.6~m \emph{Curtis--Schmidt} telescope from the University of Michigan/Cerro
Tololo Inter-American Observatory (CTIO), in three narrow-band filters
([\ion{S}{ii}]$\lambda\lambda$6716,\,6731 \AA, H$\alpha$, and
[\ion{O}{iii}]$\lambda$5007 \AA), and matching red and green continuum filters.
We also included data obtained with the MOSAIC\,{\sc ii} camera on the Blanco
\mbox{4-m} telescope at the CTIO. Although only a H$\alpha$ image is available,
the superior angular resolution of the MOSAIC\,{\sc ii} data (pixel size of
less than 1\arcsec) allow faint filaments to appear sharper.

\subsection{Infrared}
\label{observations_infrared}
To study possible emission from our objects, and study their surrounding cold
environments traced by infrared emission, we used data from the SAGE survey
\citep{2006AJ....132.2268M}, performed with the \emph{Spitzer Space
Telescope}. We essentially used data from the Multiband Imaging Photometer
\citep[MIPS;][]{2004ApJS..154...25R} in its 24 \textmu m band. We retrieved the
MIPS mosaiced, flux-calibrated (in units of MJy\,sr$^{-1}$) images processed by
the SAGE team, which have a pixel size of 2.49\arcsec.

\subsection{Radio}
\label{observations_radio}
To investigate radio emission from the sources, we used data from various radio
surveys, particularly the 4800 MHz survey by \citet{2010AJ....140.1567D} and the
1370 MHz survey by \citet{2007MNRAS.382..543H}. Both surveys used the Australia
Telescope Compact Array (ATCA) in fairly compact configurations to produce
half-power beamwidths of 35\arcsec\ and 45\arcsec, respectively.  Data at both
frequencies from a survey using the 64-m Parkes telescope
\citep{1991A&A...252..475H,1995A&AS..111..311F} were included in the imaging to
improve the sensitivity to the smooth emission from these extended regions.  We
note that none of these objects was visible on the SUMMS radio survey of the
southern sky at 843 MHz \citep[][2008 version of the data]{2003MNRAS.342.1117M}.

\begin{figure}[t]
    \begin{center}
    \includegraphics[width=1.00\hsize]{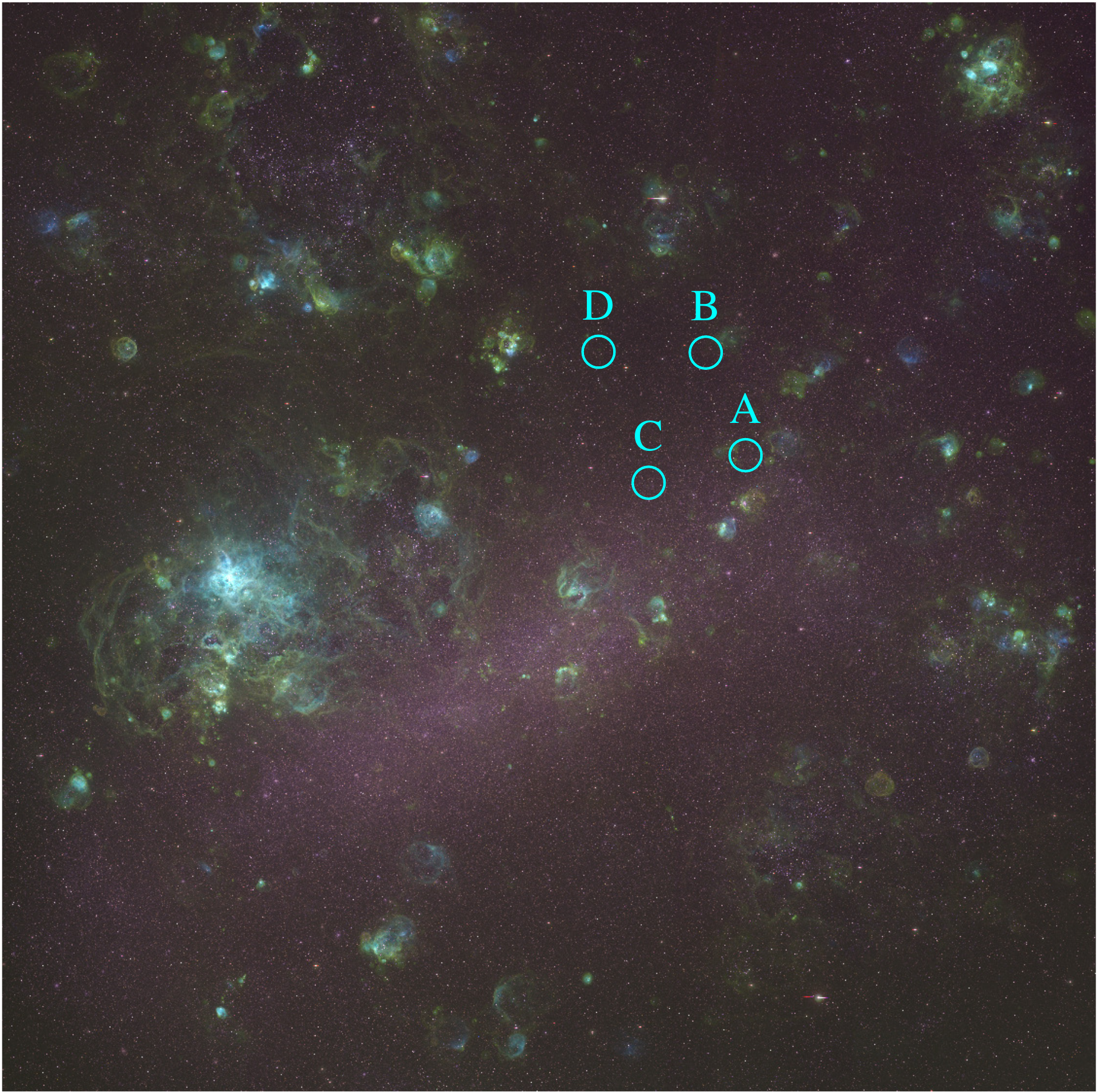}
    \end{center}
\caption{View of the LMC from MCELS data (red for [\ion{S}{ii}], green for
H$\alpha$, blue for [\ion{O}{iii}]), with positions of the new SNRs of this work
shown by the cyan circles. The letters designate the individual remnants, as
given in Table\,\ref{table_xray_observations}. The image spans 5.6\degr .}
    \label{fig_lmc_snrs}
\end{figure}

\section{Sample selection}
\label{sample_selection}
To identify SNR candidates, we first selected all extended X-ray sources from
the source detection lists. Then, we checked the mosaic images created by our
pipeline to screen out spurious sources (for instance, sources at the outer
edge of the detector). The final and most important filtering was to exclude
real sources that were not SNRs. We list below extended sources that can be
found in the field of the LMC and their main X-ray properties.

\begin{enumerate}
    \item Superbubbles (SBs) are large diameter ($\sim 100$ pc, corresponding to
$\sim 6.9$ \arcmin\ in the LMC), optically bright shells filled with hot X-ray
emitting gas. These objects are powered by the combined stellar winds of massive
stars within the shell, with possible contributions by interior SN explosions.
They are quite common in the LMC. For details and references, see e.g.
\citet{1990ApJ...365..510C} and \citet{2001ApJS..136..119D}.
    \item Galaxy clusters behind the LMC can be observed in our survey, like in
the SMC \citep{2012A&A...545A.128H}. The hot intra-cluster medium produces
thermal X-ray emission, with typical temperatures
of 2 to 10 keV \citep{2002ARA&A..40..539R}. This is much hotter than the plasma
temperatures found in middle-aged and mature SNRs ($\lesssim 1$ keV). Galaxy
cluster candidates can therefore be recognised by their distinctive hardness
ratios (HRs), the ratios of X-ray counts in various energy bands
\citep[][]{2013arXiv1307.7594S}.
    \item Hot ISM in the LMC produces diffuse emission on large scales. Knots of
higher surface brightness can be misidentified as extended sources, but visual
inspection of the mosaic images allows to find such cases easily. Study of the
hot gas in the LMC with unprecedented spatial resolution will form the core of
a future paper.
\end{enumerate}

In the available observations we identified four sources that \emph{i)} did
not belong to any of the above categories; \emph{ii)} had X-ray properties
(e.g. size and soft emission) typical of SNRs; and \emph{iii)} had enough counts
to allow a meaningful spectral analysis. Other candidates, either fainter or in
a more complex region (namely, near 30 Doradus), will be presented in future
works. A further reason why we focused first on these four sources is that
they were observed earlier, in the first pointings of the LMC survey (which is
spread over several \xmm\ Announcements of Opportunity [AO]). The location of
the sources in the LMC is shown in Fig.\,\ref{fig_lmc_snrs}.


To designate the sources we follow the Magellanic Cloud Supernova
Remnant\,\footnote{MCSNR, \url{http://www.mcsnr.org/Default.aspx}} Database
nomenclature, which uses identifiers of the form ``MCSNR JHHMM$-$DDMM''. Based
on their positions, we assign to the four objects the identifiers \object{MCSNR
J0508$-$6830}, \object{MCSNR J0511$-$6759}, \object{MCSNR J0514$-$6840}, and
\object{MCSNR J0517$-$6759}. We chose to introduce the names of the objects
here to allow an easier description of the analysis. For the same reason we will
here and after simply call them ``remnants'' (and no longer ``candidates'').
Firm evidence for their classification as SNR is presented in
Sect.\,\ref{results}.

\section{Data analysis}
\label{data}

\subsection{X-ray imaging}
\label{data_xray_imaging}
To study the size and morphology of diffuse emission and to look for
spatially-resolved spectral variations, we created images and exposure maps in
various energy bands from the filtered event lists. We selected single and
double-pixel events (\texttt{PATTERN} = 0 to 4) from the pn detector. Below 0.5
keV, however, only single-pixel events were selected to avoid the higher
detector noise contribution from the double-pixel events. From the MOS detectors
we selected all single to quadruple-pixel (\texttt{PATTERN} = 0 to 12) events.
Masks were applied to filter out bad pixels and columns.

We used three energy bands suited to the analysis of the (mostly thermal)
spectra of SNRs. A soft band from 0.3 to 0.7 keV includes strong lines from
oxygen; a medium band from 0.7 to 1.1 keV comprises Fe L-shell lines as well as
Ne He$\alpha$ and Ly$\alpha$ lines; and a hard band (1.1--4.2 keV) which
includes lines from Mg, Si, S, Ca, Ar, and possibly non-thermal continuum.

We used filter-wheel closed (FWC) data, obtained with the detector shielded
from the astrophysical background, to subtract the detector background. We
estimated the contribution of detector background in each observation from the
count rates in the corner of the images, which were not exposed to the sky. We
then subtracted approriately-scaled FWC data from the raw images.

Images from pn and MOS detectors were merged together. In the case of \snrc,
two observations included the source in the field of view, and we merged images
from the two datasets together. Next, we performed adaptive smoothing: the sizes
of Gaussian kernels were computed at each position in order to reach a
signal-to-noise ratio of five, setting the minimum full width at half maximum
of the kernels to 20\arcsec. In the end the smoothed images were divided by the
corresponding vignetted exposure maps.

With the smoothed, exposure-corrected, and detector background-subtracted
images in three bands, we created X-ray ``true colour'' images of the four
objects in our sample. These are shown in Figs.\,\ref{fig_rgb_image_snra} to
\ref{fig_rgb_image_snrd}. The analysis of the morphology and size of each
remnant is presented in Sect.\,\ref{results_morphology}.

\subsection{X-ray spectral analysis}
\label{data_xray_spectral}
The effective area of the X-ray telescopes varies with off-axis angle across the
source extent. To take that into account we extracted spectra from
vignetting-corrected event lists, produced with the SAS task
\texttt{evigweight}. Only EPIC-pn spectra were used for the spectral analysis
because all four remnants are very faint in the MOS-only images.

Except for \snrd, the extraction regions were circles including all the X-ray
emission. Because of the roughly triangular morphology of \snrd\ (see
Sect.\,\ref{results_morphology_snrd}), we manually defined a polygonal region
following the remnant's X-ray emission. We extracted background spectra from
adjacent regions, which were chosen to be at a similar off-axis angle and on the
same CCD chip as the source. However, we note that all remnants but \snrb\ were
located on more than one chip, possibly leading to systematic uncertainties in
computing response functions and subtracting the background. Point sources
detected in the extraction regions were excised.

EPIC-pn spectra including single and double-pixel events were rebinned using
the FTOOL \texttt{grppha} to have a minimum of 25 counts per bin. This allows
the use of the $\chi ^2$-statistic in XSPEC \citep{1996ASPC..101...17A} version
12.8.0, which we utilised for the spectral analysis.

\subsubsection{Background modeling}
\label{data_xray_spectral_background}
The spectral analysis of extended sources with low surface brightness requires
a careful treatment of the background, both instrumental and astrophysical. It
is not desirable to simply subtract a background spectrum extracted from a
nearby region, because of the different responses and background contributions
associated to different regions, and because of the resulting loss in the
statistical quality of the source spectrum. Instead, we explicitly model the
background and simultaneously fit the source and background spectra. This
technique was already applied in our recent analyses of new LMC SNRs 
\citep{2012A&A...546A.109M,bozzetto2013}.

The instrumental background \citep{2002A&A...389...93L,2004SPIE.5165..112F}
comprises
\emph{i)} a particle-induced component called Quiescent Particle Background
(QPB), from the interaction of high-energy particles with the detectors,
\emph{ii)} fluorescence lines emitted by the material in the cameras, and
\emph{iii)} internal electronic noise (dominating below 400 eV).

We used FWC data (Sect.\,\ref{data_xray_imaging}) to model the instrumental
background, as no astrophysical X-ray photons are present in these data. Spectra
were extracted from FWC data at the same positions on the detector as the source
and background extraction regions for each SNR.
We applied the same screening and filtering criteria used for the science (i.e.
observational) data to the FWC data, and we used the task \texttt{evigweight} to
apply a vignetting correction. Formally, the instrumental background is
\emph{not} subjected to vignetting, which is an effect of the telescope on
\emph{photons}. However, by applying a vignetting correction to the
science data, we assign weights to genuine X-ray events as well as to particle
background events, since we cannot \emph{a priori} distinguish the two type of
events. Therefore, we need to vignetting-correct the FWC data to make sure that
the FWC spectra, used for the modeling of the instrumental background
contribution to the science data, have been processed in the same way as the
latter. At a given position on the detector, \texttt{evigweight} will assign
heavier weights to photons with higher energies, an effect that can be easily
accounted for in our background model.

The empirical model developed by \citet{2012PhDT......ppppS} was used to
describe the FWC spectra. This includes an exponential decay (modified by a
spline function), a power law, and a combination of Gaussian lines to account
for the electronic noise, QPB, and instrumental lines, respectively. In
addition, two smeared absorption edges to the continuum are included. Finally,
we added a spline function to model the effect of the vignetting correction,
which ``overweights'' events above 5 keV, if they have been recorded at
significant off-axis angles. This model was not convolved to the instrumental
response and we used a normalised diagonal response in XSPEC.

The astrophysical background was modeled by the three-component model of
\citet[][their Eq.\,2]{2010ApJS..188...46K}. The temperature of the unabsorbed
thermal component associated to the Local Hot Bubble or Solar Wind Charge
Exchange emission was fixed to 100 eV \citep{2008ApJ...676..335H}. The photon
index of the power law used to model the unresolved extragalactic background was
fixed to 1.46 \citep{1997MNRAS.285..449C}, whilst the temperature of the
(absorbed) Galactic halo emission was allowed to vary.

There could also be X-ray (thermal) emission from hot gas in the LMC that should
be taken into account to extract the purest information on the SNRs. This does
not occur in any case presented here: First, the four SNRs are in a region of
the LMC devoid of significant diffuse emission; such emission is seen more
south-east of the remnants, close to the LMC bar and around 30 Doradus, showing
where most of the LMC hot gas is located. Second, this diffuse emission would be
seen as a thermal component (with temperatures $kT \sim$ 0.2--0.3 keV) and would
not be spectrally resolved from the Galactic halo emission. Therefore, any
LMC-intrinsic background emission would be accounted for by the hot thermal
emission in the three-component background model used.

 Finally, to account for a
possible soft proton contamination (SPC) in the science spectra, we included an
additional power law component, not convolved with the instrumental response.

\subsubsection{Source emission}
\label{data_xray_spectral_source}
To fit the X-ray emission from the remnants, we first used single-temperature
models, assuming either collisional ionisation equilibrium (CIE) or
non-equilibrium ionisation (NEI). The CIE model used is \emph{vapec} in XSPEC
and makes use of the Astrophysical Plasma Emission Code (APEC). We used the most
recent version available (v2.0.2), which includes updated atomic data
\citep{2012ApJ...756..128F}. The update of the collisional excitation data of
many iron L-shell ions is particularly relevant to our analysis, as we shall
see in Sect.\,\ref{results_spectroscopy}. As NEI model we chose the
plane-parallel shock model of \citet[][\emph{vpshock} in
XSPEC]{2001ApJ...548..820B}. It features a linear distribution of ionisation
ages ($\tau = n_e t$, where $n_e$ is the electron density and $t$ the time since
the plasma was shocked) which is more realistic than the single-ionisation age
often used in analysis of X-ray spectra of SNRs.

The data did not require\,/\,allow the use of multi-temperature models. For the
spectrum of \snrc, which has the best photon statistics due to its higher
surface brightness\,/\,extent and the fact that it is covered in two \xmm\
observations, we also tried a Sedov model (\emph{vsedov} in XSPEC; the prefix
``v'' in the XSPEC names indicates that abundances can vary). This Sedov model
\citep{2001ApJ...548..820B} computes the X-ray spectrum of an SNR in the
Sedov-Taylor stage of its evolution.

For the Galactic foreground absorption we used a photoelectric absorption
model (\emph{phabs} in XSPEC) at solar metallicity, with cross-sections taken
from \citet{1992ApJ...400..699B}. We included an additional absorption column
to the SNR components to model absorption by atomic gas in the LMC. The
abundance patterns of these components were fixed to that of the LMC, as
measured by \citet{1992ApJ...384..508R}, and the columns $N_{H\ \mathrm{LMC}}$
were a free parameter in the fits. Elemental abundances of the SNR components
were also initially fixed to those of \citet{1992ApJ...384..508R}. We then
determined whether the data required different abundance patterns (see the
individual description in Sect.\,\ref{results_spectroscopy}). All abundances are
given relative to the solar values as given by \citet{2000ApJ...542..914W}.

We simultaneously fit the FWC and science data extracted from the background and
source regions. The detector background model constrained by the FWC data is
included in the corresponding science spectrum, fixing the parameters and
allowing only a constant factor between the FWC and science spectra. Likewise,
the parameters of the astrophysical background and SPC components in the SNR
spectra were linked to the corresponding parameters constrained by the science
background spectra.

\subsection{Optical}
\label{Optical}
We combined all MCELS observations covering the four SNRs and smoothed the
flux-calibrated images with a 2\arcsec\ Gaussian. We subtracted the
corresponding continuum images , thereby removing (most of) the stellar
contribution and revealing the faint diffuse emission in its full extent.

\begin{figure*}[t]
    \includegraphics[width=0.49\hsize]
{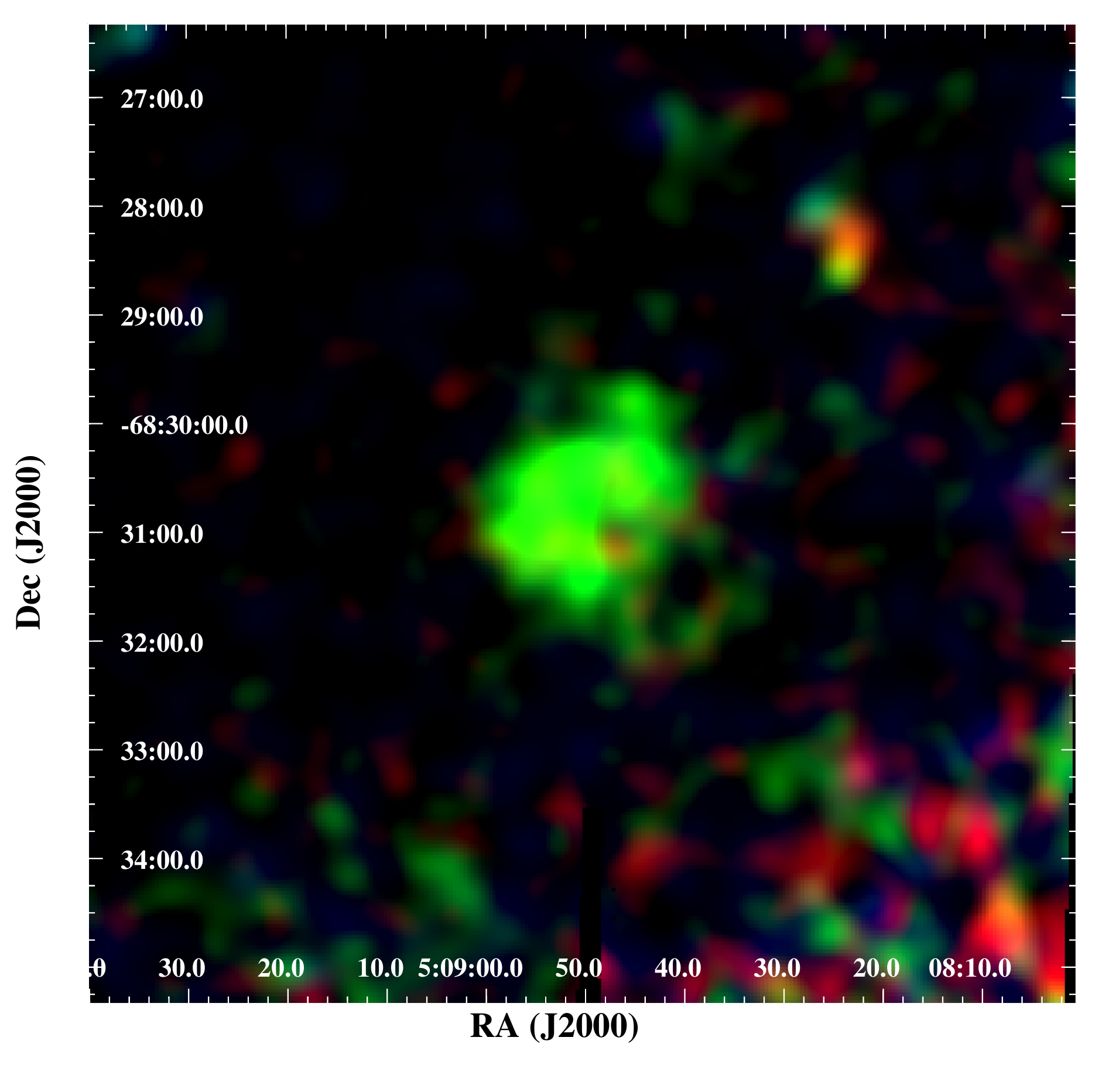}
    \includegraphics[width=0.49\hsize]
{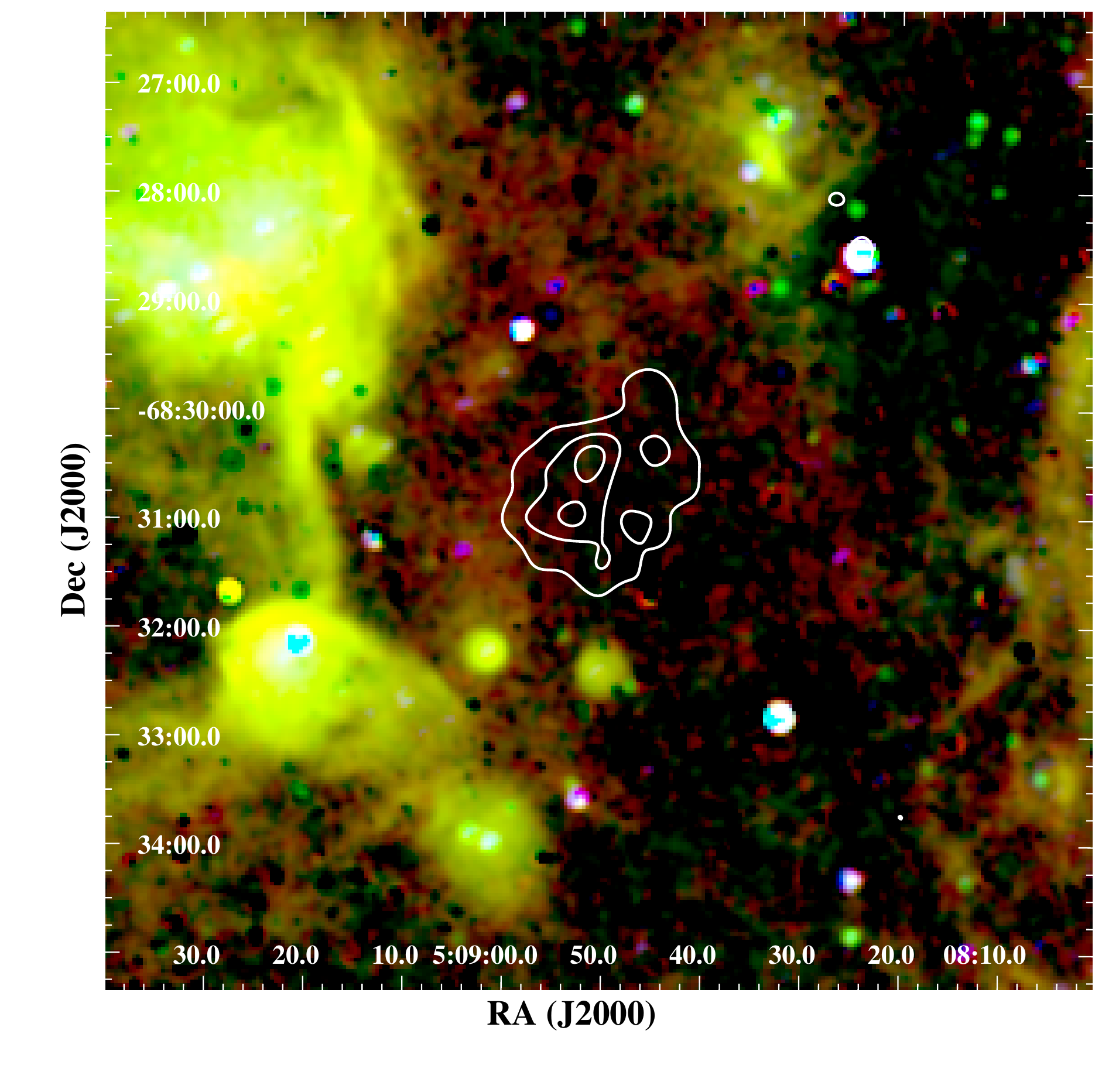}

    \includegraphics[width=0.49\hsize]
{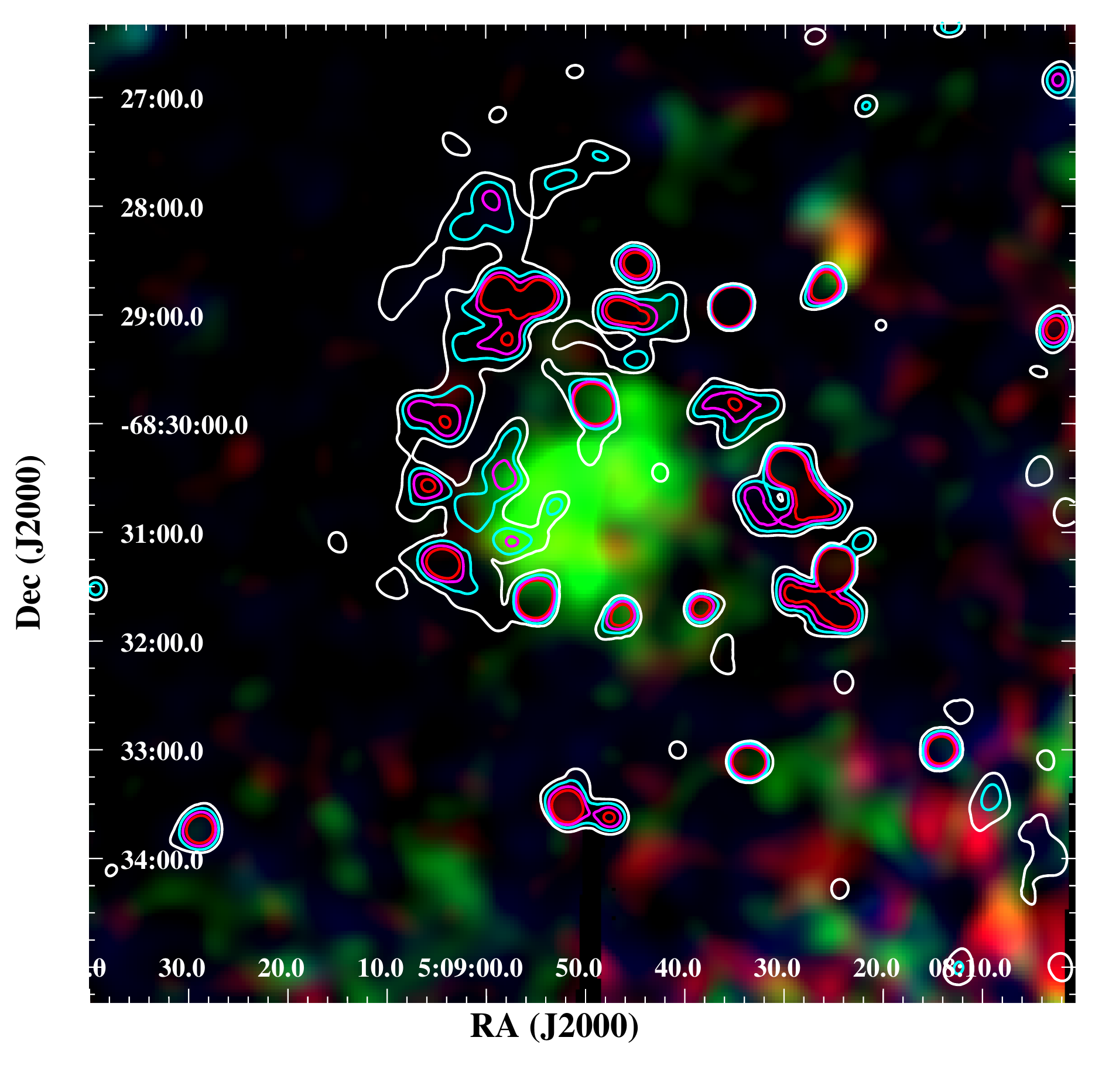}
    \includegraphics[width=0.49\hsize,viewport= -20 -40 565 545, clip]
    {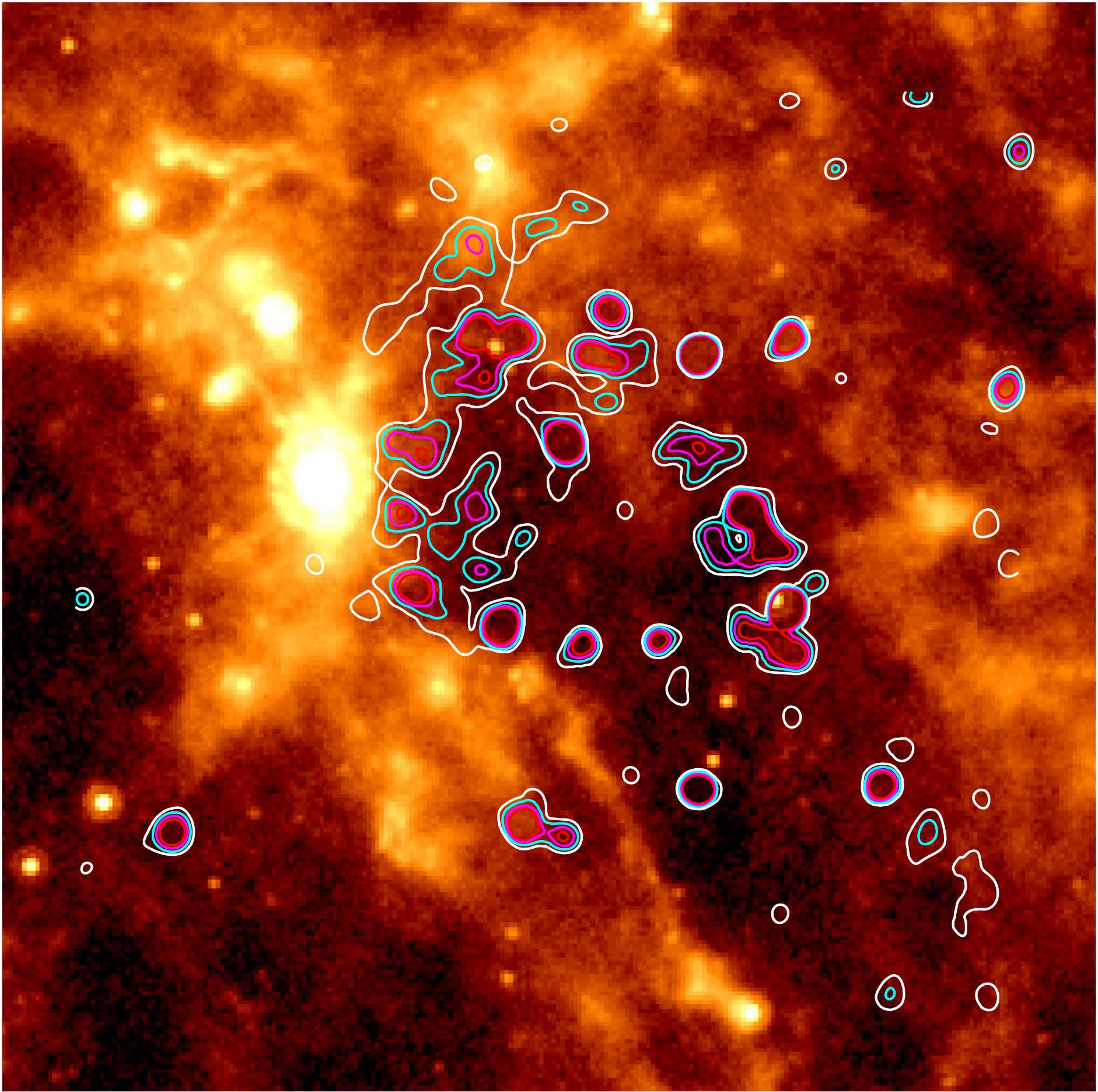}
\sidecaption
\caption{A multicolour view of \snra.
\emph{Top left}\,: X-ray colour image of the remnant, combining the data from
all EPIC cameras. The red, green, and blue components are soft, medium, and hard
X-rays, as defined in the Sect.\,\ref{data_xray_imaging}.
\emph{Top right}\,: The same region of the sky in the light of [\ion{S}{ii}]
(red), H$\alpha$ (green), and [\ion{O}{iii}] (blue),where all data are from the
MCELS. The X-ray contours from the medium band are overlaid.
\emph{Bottom left}\,: Same EPIC image as above but with
[\ion{S}{ii}]-to-H$\alpha$ ratio contours from MCELS data. Levels are
at 0.6, 0.8, 1.0, and 1.2 in white, cyan, magenta, and red, respectively
\emph{Bottom right}\,: The remnant as seen at 24 \textmu m, with the same
[\ion{S}{ii}]-to-H$\alpha$ ratio contours as on the left.
}
    \label{fig_rgb_image_snra}
\end{figure*}

\begin{figure*}[t]
    \includegraphics[width=0.49\hsize]{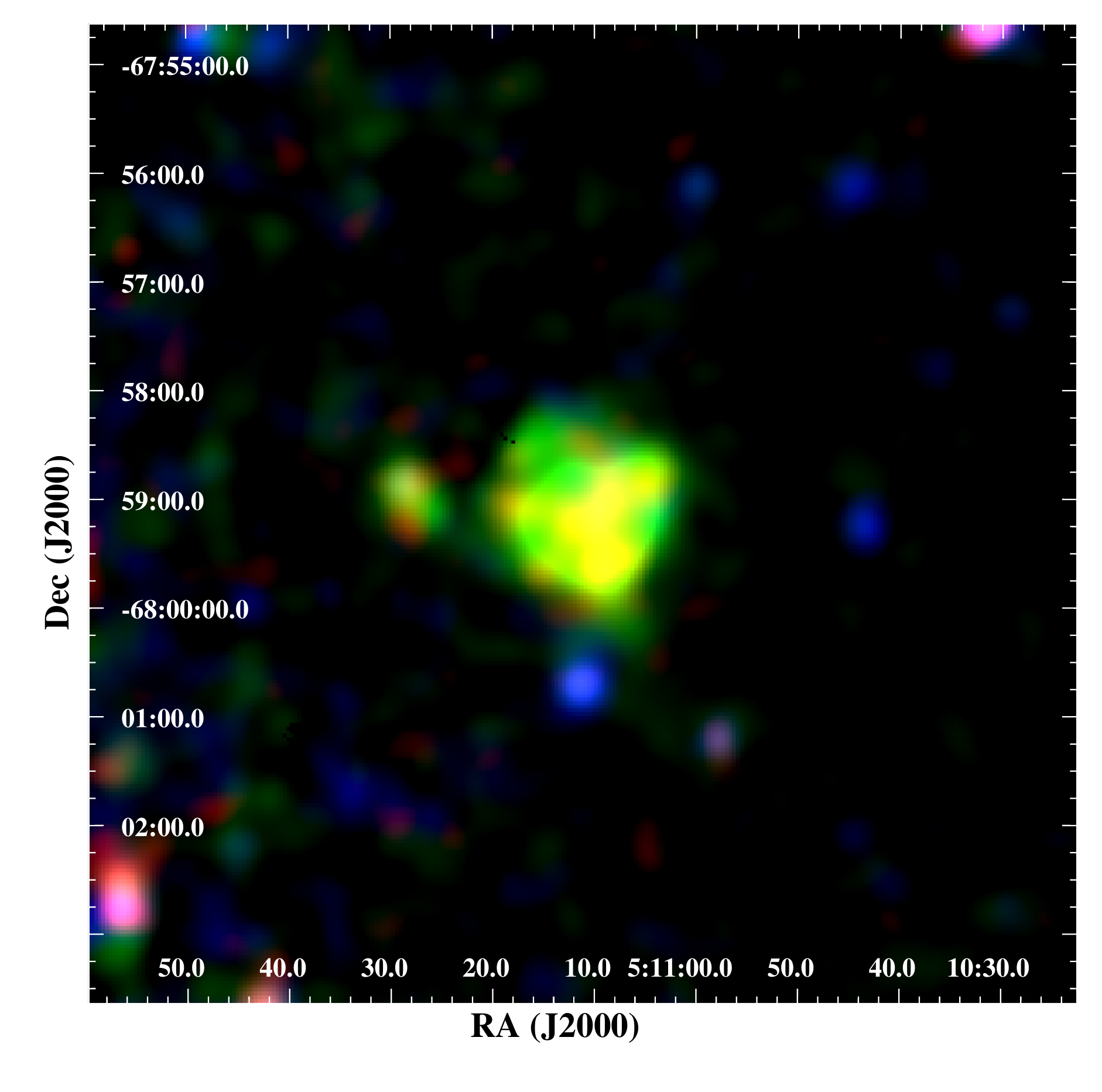}
    \includegraphics[width=0.49\hsize]{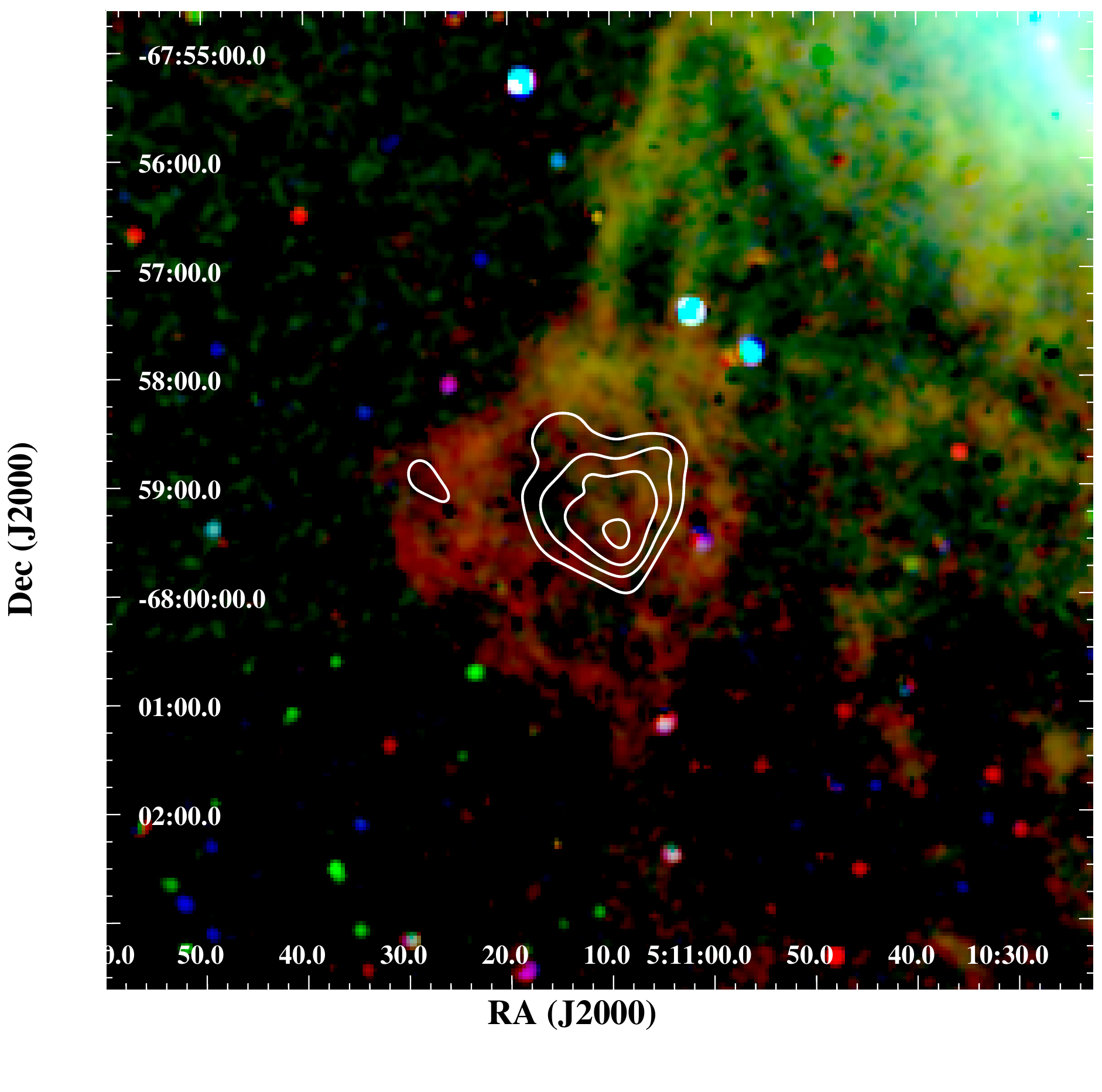}

    \includegraphics[width=0.49\hsize]
    {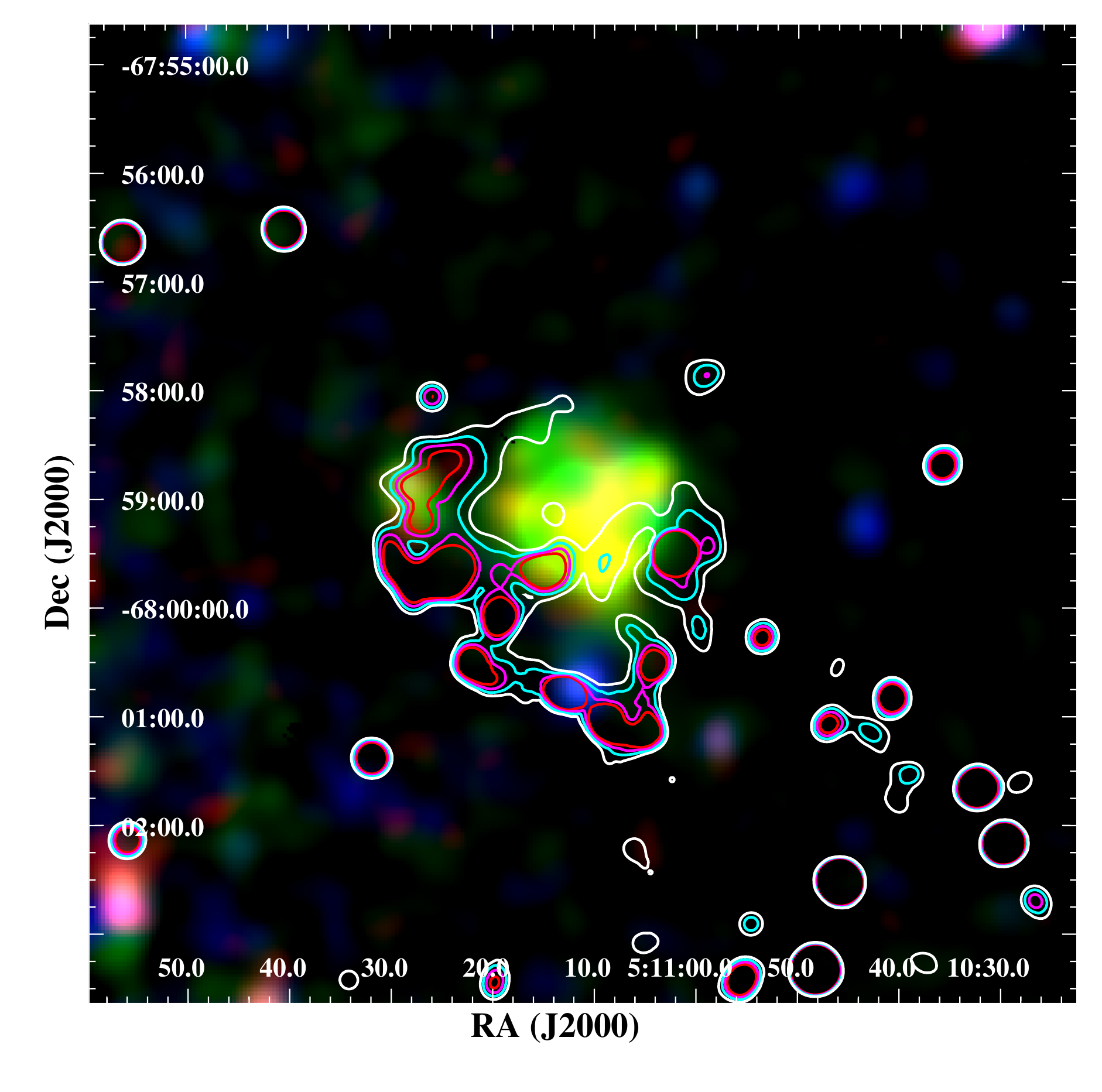}
    \includegraphics[width=0.49\hsize,viewport= -30 -40 545 545, clip]
{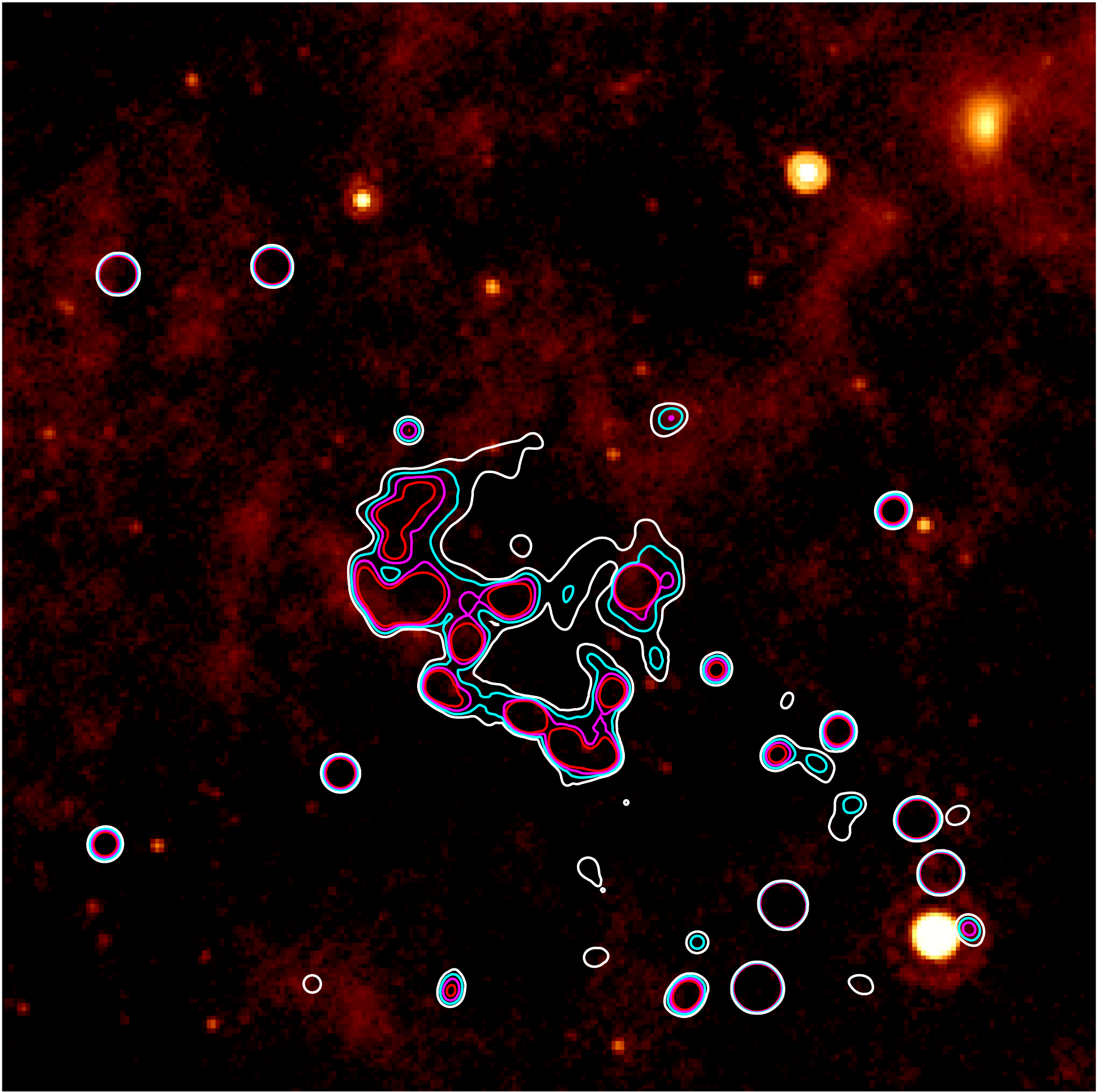}

\caption{Same as Fig.\,\ref{fig_rgb_image_snra} for \snrb.}
    \label{fig_rgb_image_snrb}
\end{figure*}

\begin{figure*}[t]
    \includegraphics[width=0.49\hsize]{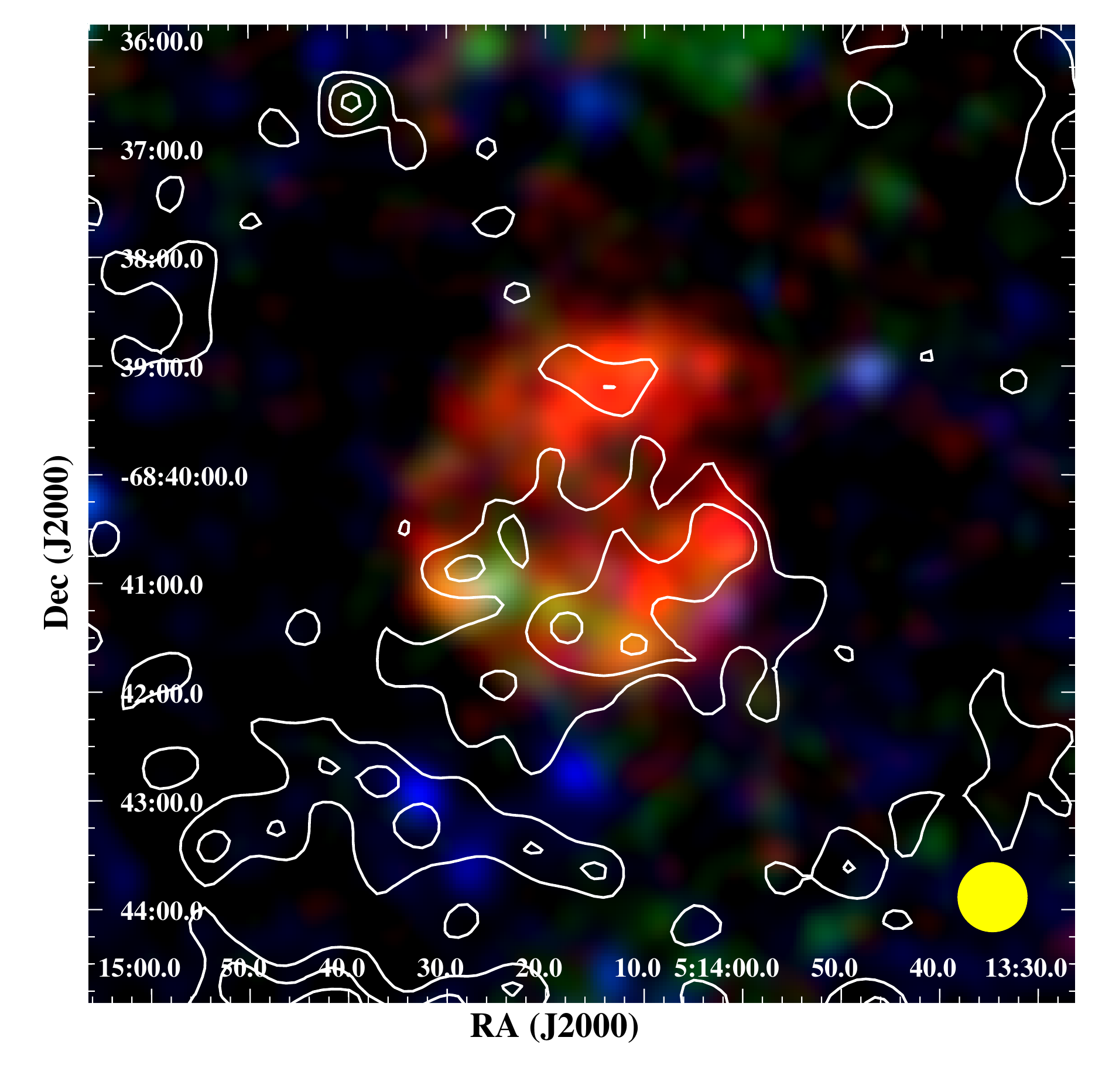}
    \includegraphics[width=0.49\hsize]{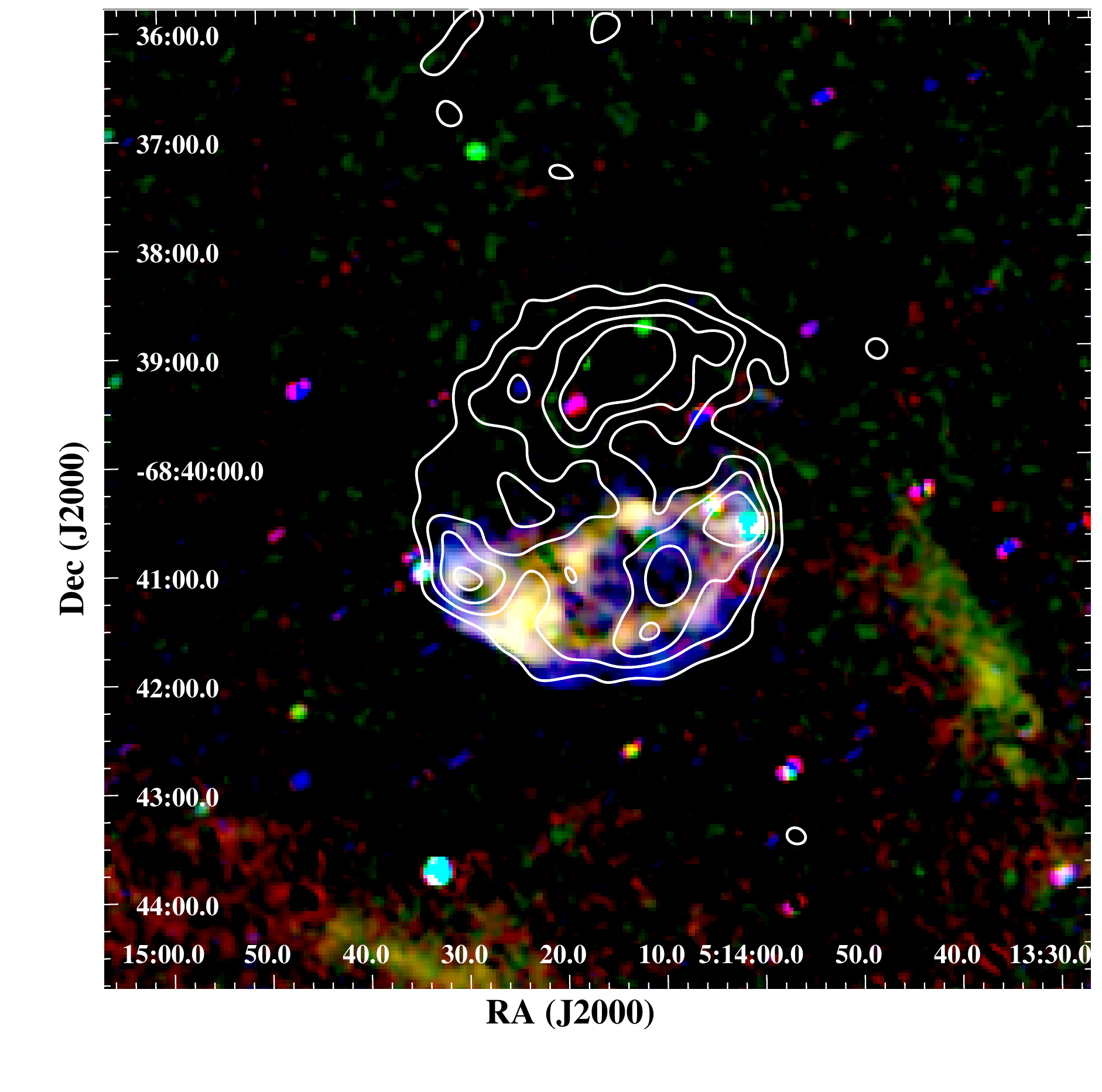}

    \includegraphics[width=0.49\hsize]
    {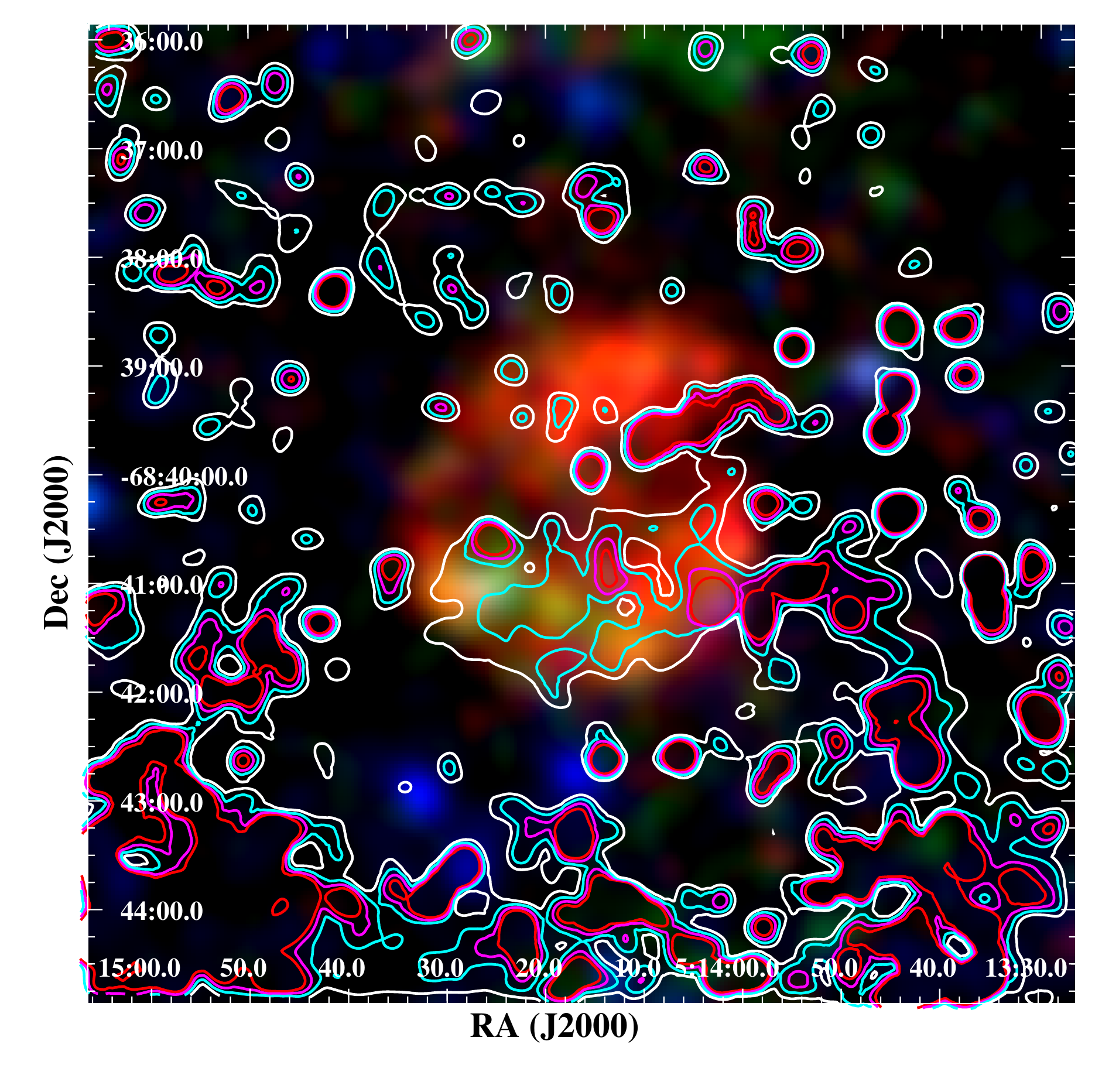}
    \includegraphics[width=0.49\hsize,viewport= -20 -40 565 545, clip]
{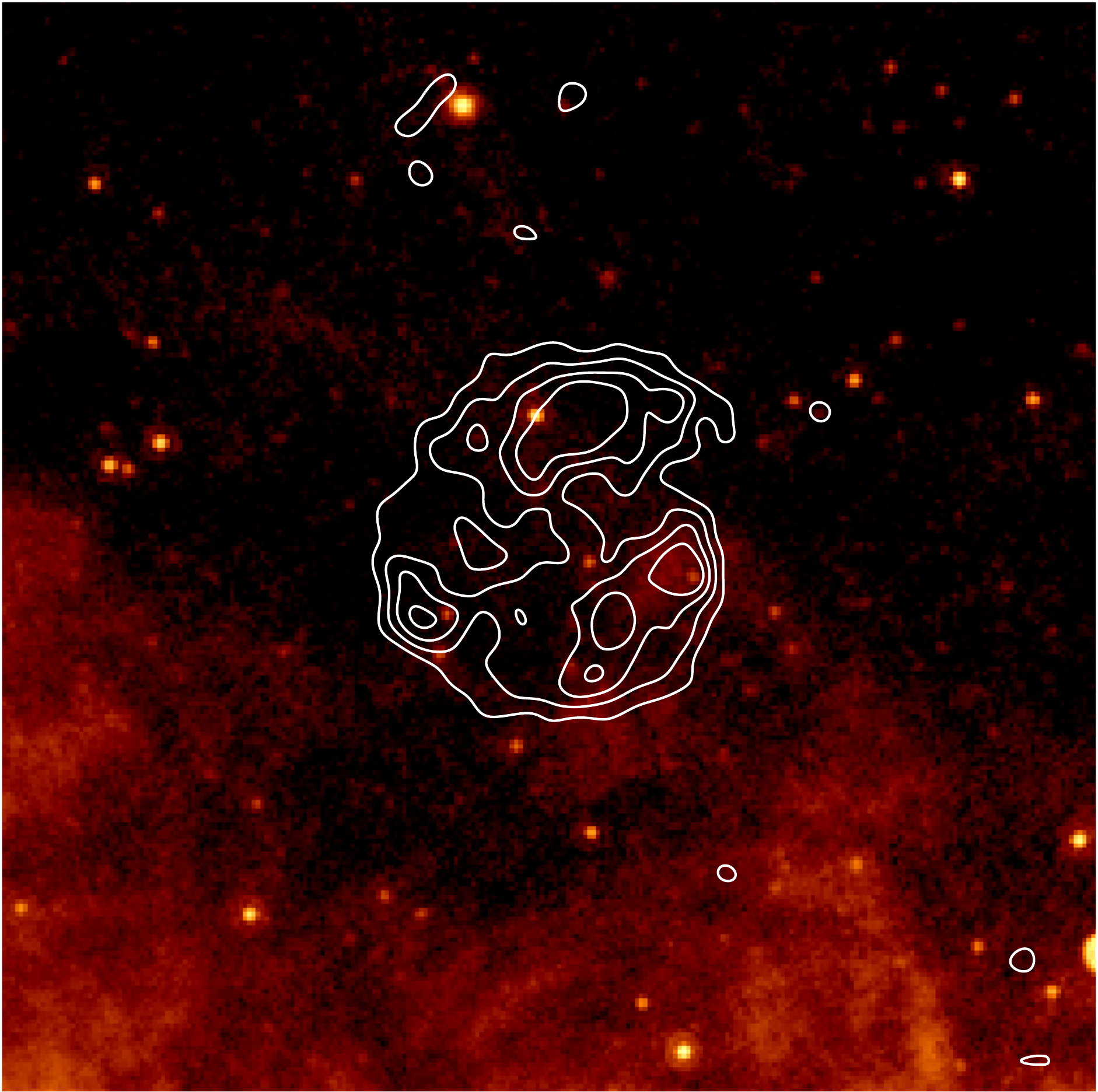}
\sidecaption
\caption{Same as Fig.\,\ref{fig_rgb_image_snra} for \snrc. On the \xmm\ image
(top left) the 4800 MHz contours are shown in white. Levels are at 1.5, 2, 2.5,
and 3 mJy\,beam$^{-1}$. The yellow disc in the lower right corner indicates the
half-power beamwidth of 35\arcsec. The X-ray contours used on the optical image
(top right) are from the soft X-ray image. On the MIPS image (bottom right) the
X-ray contours are used to locate the position of the remnant, rather than the
(noisy) \siiha\ ratio contours.}
    \label{fig_rgb_image_snrc}
\end{figure*}

\begin{figure*}[t]
    \includegraphics[width=0.49\hsize]{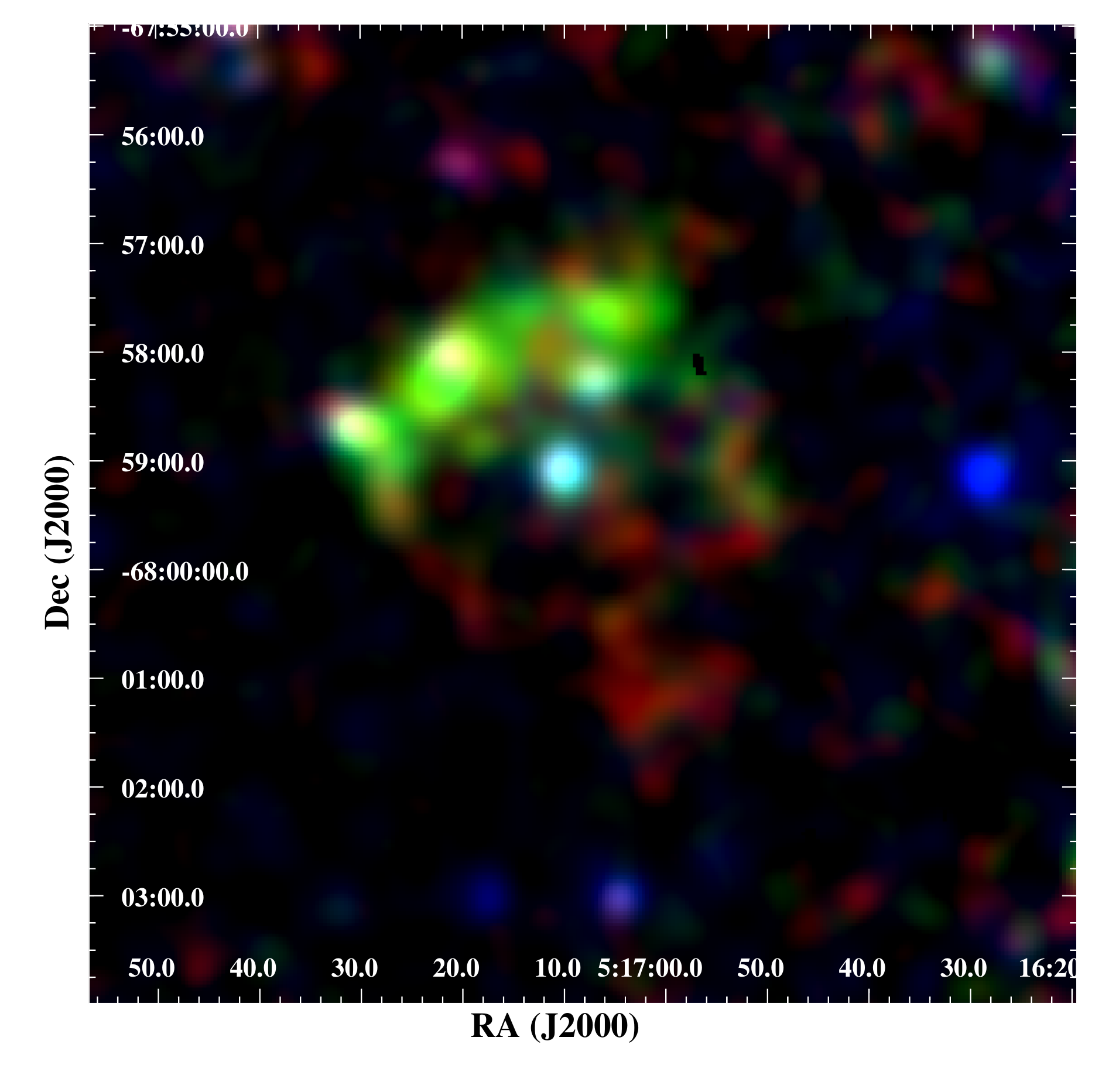}
    \includegraphics[width=0.49\hsize]{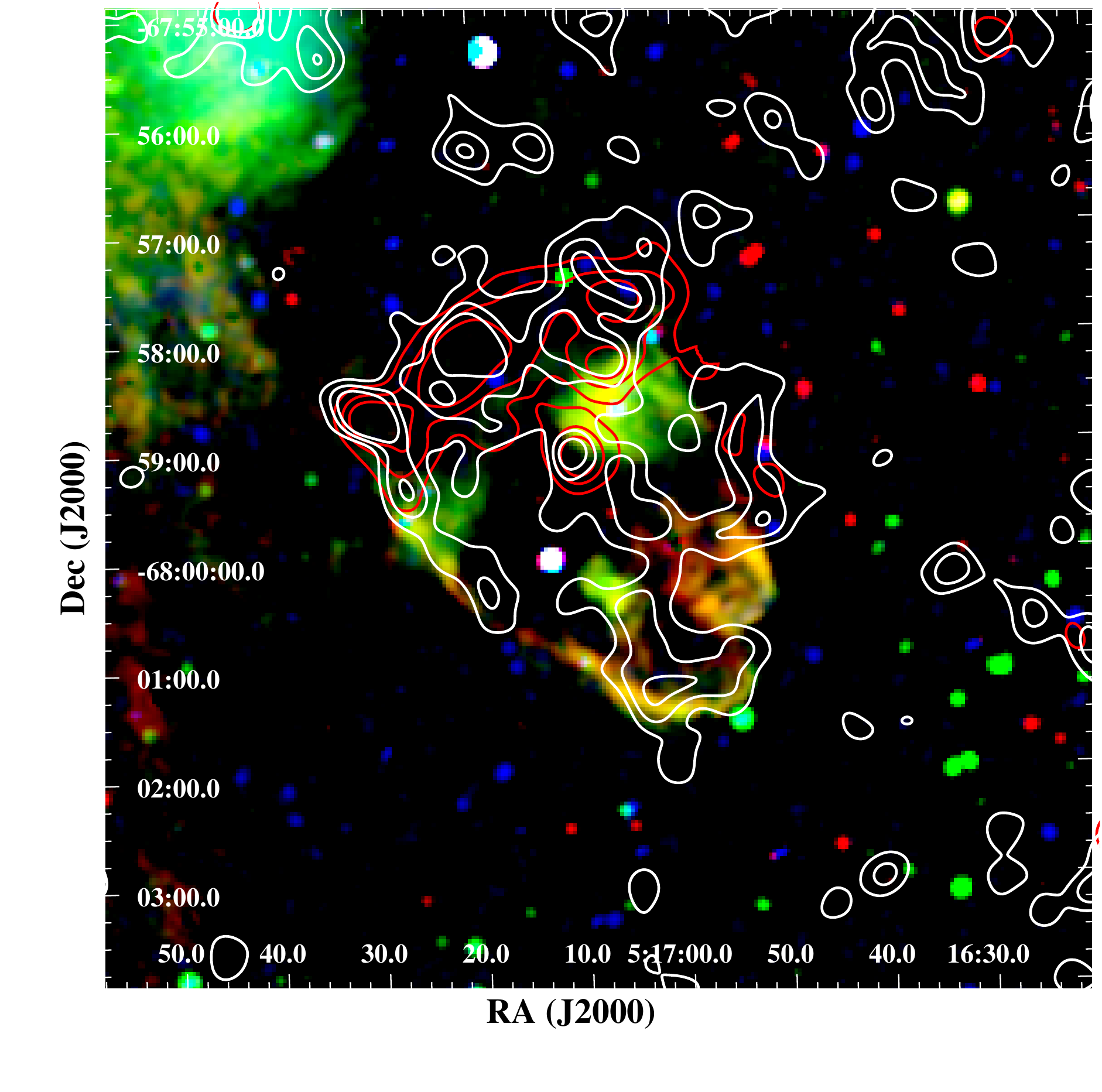}

    \includegraphics[width=0.49\hsize]
        {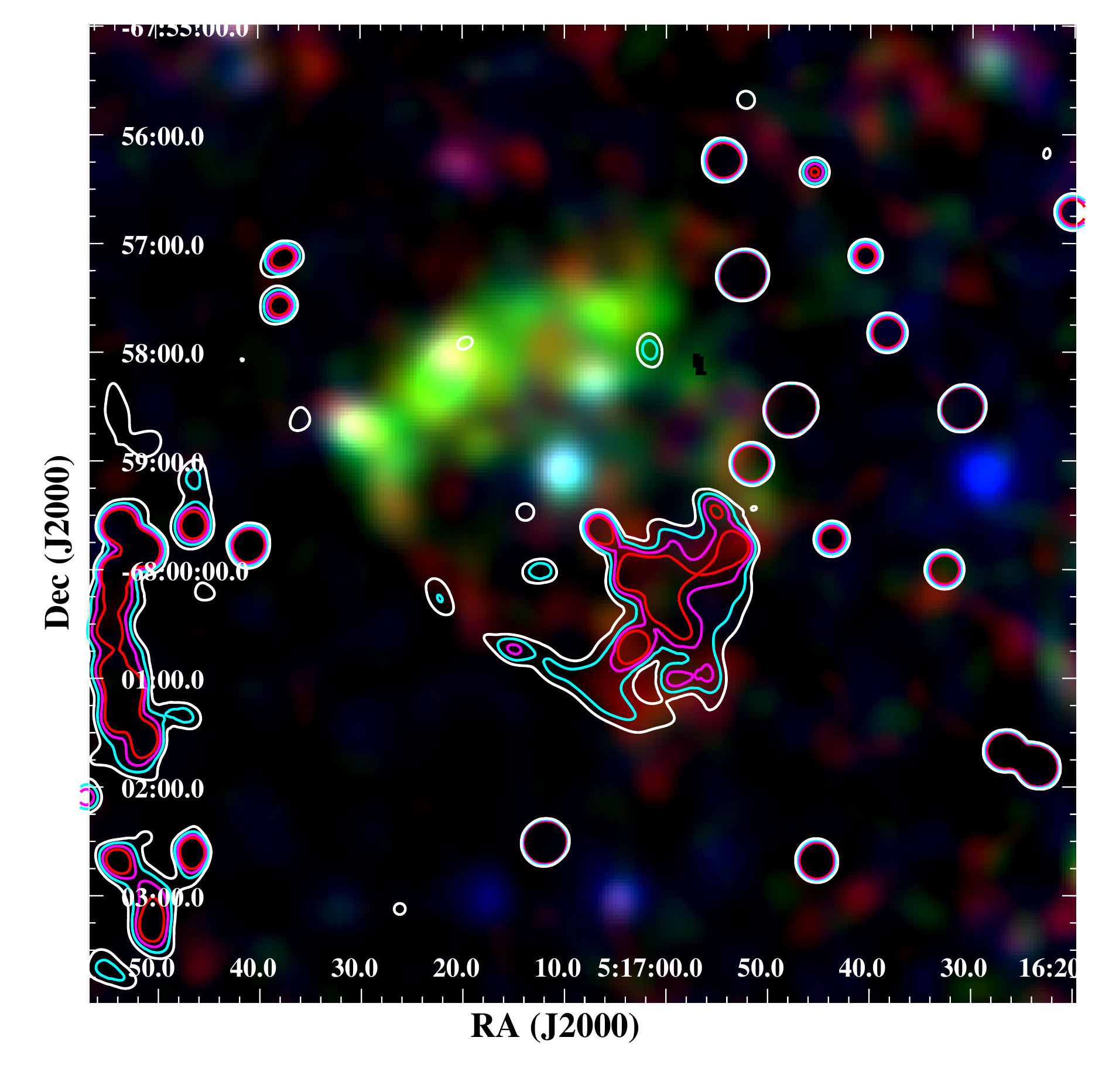}
    \includegraphics[width=0.49\hsize,viewport= -30 -40 545 545, clip]
{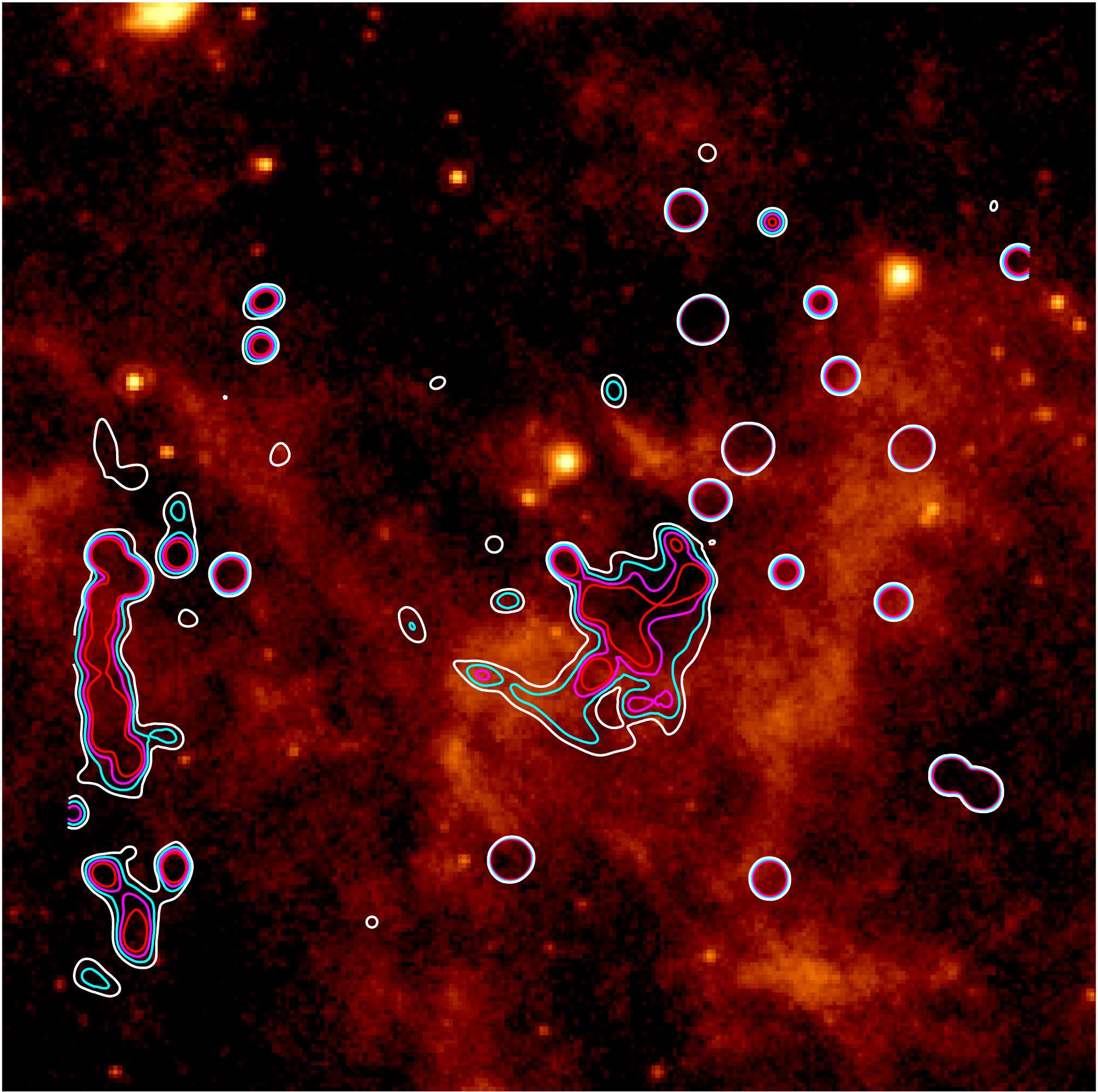}
\caption{Same as Fig.\,\ref{fig_rgb_image_snra} for \snrd. On the optical image
(top right) we show the soft band X-ray contours in white and the medium band
X-ray contours in red. }
    \label{fig_rgb_image_snrd}
\end{figure*}

A [\ion{S}{ii}]/H$\alpha$ ratio map was produced. To avoid noise where the pixel
values in either bands were low or negative (due to over-subtraction of the
continuum, particularly around stars), we set the ratio to 0 for these pixels.
From this map we could investigate possible strong \siiha\ ratios, which are
indicative of shock excitation. (In the case of photoionisation by a stellar UV
field, sulphur is brought to higher ionisation stages, reducing the \siiha\
ratio.) A classical threshold is \siiha\ $>$ 0.4 \citep{1973ApJ...180..725M},
but see discussion and references in \citet{1985ApJ...292...29F}. We used a
more conservative criterion of 0.6 to identify regions where the ratio is
clearly enhanced. \siiha\ contours around the SNRs are shown in
Figs.\,\ref{fig_rgb_image_snra} to \ref{fig_rgb_image_snrd}, and the results
described in Sect.\,\ref{results_morphology}.

\section{Results}
\label{results}

\subsection{Multi-wavelength morphology}
\label{results_morphology}
We discuss below the morphology of each remnant. We supplement the dominating
X-ray emission with data at longer wavelengths. The sizes and centres of the
SNRs (all given in J2000 equinox) are listed in
Table\,\ref{table_xray_observations}.

\label{results_morphology_snra}
\paragraph{\snra:} This remnant is the second faintest source in our sample. It
emits X-rays chiefly in the medium band (0.7--1.1 keV), with little or no flux
in the softer and harder bands, respectively. This suggests the predominance of
iron emission and is explored in greater detail in our spectral analysis
(Sect.\,\ref{results_spectroscopy}).

To estimate the size and nominal centre of the X-ray remnant we fit an ellipse
to the 0.7--1.1 keV contours. The outer contour was chosen to be 0.26 times the
amplitude (peak intensity minus average background intensity). This would
enclose 90 \%\ of the flux of a Gaussian-distributed profile. Size
uncertainties were estimated by changing the outer contours by $\pm 10 \%$ of
the total amplitude and then re-computing the size of the ellipse.
We found a nominal centre of RA = 05\hour\,08\minute\,49.5\second, DEC =
$-$68\degr\,30\minute\,41\second, a semi-major and semi-minor axis of
1.15\arcmin\ and 0.90\arcmin, respectively, and a position angle (PA) of
40\degr\ west of north. The uncertainty is 0.1\arcmin\ in each direction. At
the assumed distance of the LMC, this gives an extent of \snra\ of $16.7 (\pm
1.5)$ pc $\times 13.1 (\pm 1.5)$ pc.

Whilst there is no obvious association in the MCELS images to the X-ray
emission, we do detect very faint [\ion{S}{ii}] emission, which encircles the
remnant seen in X-rays almost completely, except in the south
(Fig.\,\ref{fig_rgb_image_snra}, top right) . At the position of the
[\ion{S}{ii}] shell there is no contribution from H$\alpha$, which is only
emitted by the nearby, most likely unrelated, \ion{H}{II} regions. Therefore,
this shell and its location around the central X-ray emission appears more
clearly in the X-ray image \textit{vs.} \siiha\ contours
(Fig.\,\ref{fig_rgb_image_snra}, bottom left). The [\ion{S}{ii}] shell is twice
as large as the X-ray--emitting region (semi-axes of about
1.9$\times$2.0\arcmin). We discuss this morphology in light of the spectral
results and in comparison with other SNRs in Sect.\,\ref{discussion_comparison}.

\snra\ is not obvious on the radio images but we do see the nearby
\ion{H}{II} region to the northeast and curving around to the north and
southeast as shown in the optical images. The lack of association of radio
emission with the faint [\ion{S}{ii}] shell or bright X-ray central region
indicates the faintness of the SNR at radio frequencies and/or the insufficient
sensitivity of the current radio surveys.

In the infrared, there is no evident emission from the remnant. Most of the
diffuse emission of that region (Fig.\,\ref{fig_rgb_image_snra}, bottom right)
can be associated to the neaby \ion{H}{II} regions seen in radio and in the
optical.

\label{results_morphology_snrb}
\paragraph{\snrb:} The X-ray colour of this source is very similar to \snra, in
the sense that the 0.7--1.1 keV band totally dominates the X-ray emission. The
morphology is roughly spherical, so we adjusted a circle on the intensity map to
derive the position of the centre\,: RA = 05\hour\,11\minute\,10.7\second, DEC =
$-$67\degr\,59\minute\,07\second. To measure the size and associated
uncertainty for the source, we extracted intensity profiles intersecting the
remnant's centre, at ten different position angles. We measured the extent at
which the intensity falls below 0.26 times the amplitude (the same criterion as
for \snra\ can be applied as well for this remnant as they have similar
morphologies). We repeated this measurement for each PA, before computing the
mean and standard deviation of the ten measurements. We obtained a radius of
0.93\arcmin $\pm$ 0.09\arcmin, corresponding to a physical size of \mbox{$13.5
(\pm1.3)$ pc.}

In the continuum-subtracted [\ion{S}{ii}] images we see faint diffuse emission
at the position of \snrb. This optical emission has a roughly circular
morphology, encasing the bright X-ray emission. It appears slightly
limb-brightened, indicating a shell morphology, and its extent is
$\sim$3.8\arcmin$\times$3.6\arcmin, i.e. larger than the X-ray emission.
H$\alpha$ emission is seen at the same location, albeit even fainter, whilst
[\ion{O}{iii}]$\lambda$5007 \AA\ is completely absent. Despite the faintness of
this optical emission, its shell-like morphology and its strong \siiha\ ratio
(in excess of 0.6 and reaching 1.5) allows to secure the association of the
optical emssion to the X-rays, and to clearly discriminate the remnant from the
ambient optical emission, e.g. from the \ion{H}{II} region DEM L89
\citep{1976MmRAS..81...89D} located $\sim$8\arcmin\ to the north-west. We note
the striking similarity of the morphological and spectral features of
\snrb\ to those of \snra, and refer the reader to the discussion in
Sect.\,\ref{discussion_comparison}.

In addition, we note the presence of a small knot of X-ray emission, $\sim
1.7$\arcmin\ towards the east, \emph{outside} of the main X-ray-emitting
region. Its morphology is different from that of a point source, and its
colour\,/\,hardness ratios are very similar to that of the bulk of the X-ray
emission. Furthermore, it is located at the eastern tip of the diffuse optical
emission described above, suggesting that the knot is likely to belong to the
remnant, possibly being a clump of X-ray emitting ejecta (``schrapnel'').

\snrb\ was too faint to be detected in the various radio surveys of the LMC.
The diffuse infrared emission (at 24 \textmu m) around the remnant is very
moderate. A weak filament is found to correlate with the south-eastern part of
the [\ion{S}{ii}] shell, suggesting a physical association with \snrb. We
discuss a possible origin of this emission in
Sect.\,\ref{discussion_comparison}.

\label{results_morphology_snrc}
\paragraph{\snrc:} Unlike the two aforementioned objects, this source has softer
colours, being dominated by emission in the 0.3--0.7 keV band. Globally, the
morphology is spherically symmetric, although the southern limb is slightly
brighter than the northern one. A darker lane also appears to separate the two
halves along the east-west equator. Analysis of the X-ray spectrum provide clues
to the origin of these features (Sect.\,\ref{results_spectroscopy_snrc}).

Optical emission is present, correlating with the southern edge of the X-ray
shell (Fig.\,\ref{fig_rgb_image_snrc}, top right). Both H$\alpha$ and
[\ion{S}{ii}] lines are detected in emission: though the \siiha\ ratio map is
still noisy due to the very low diffuse emission of that region, the optical
emission at the position of the remnant is clearly in excess of 0.6, indicative
of shock--excitation. In addition, [\ion{O}{iii}]$\lambda$5007 \AA\ emission is
present, outlining the edges of the H$\alpha$ and [\ion{S}{ii}]--emitting
regions.

The position and size of the remnant were obtained from the X-ray image in the
same fashion as for \snrb. We found a centre located at RA =
05\hour\,14\minute\,15.5\second, DEC = $-$68\degr\,40\minute\,14\second, and a
radius of 1.83\arcmin $\pm 0.12$\arcmin, i.e. $26.5 (\pm1.7)$ pc.

\snrc\ is the only source of the sample clearly detected at radio frequencies.
At 4800 MHz, it looks somewhat like the optical line images with the brightest
emission on the southern side.  The X-ray image has a more circular outline. 
The 1370 MHz radio image also seems to have a rather sharp north-south gradient
across the whole image near the top of the SNR.  The integrated radio flux
densities of this SNR are quite uncertain due to the low intensity emission and
the relatively high r.m.s. noise surrounding this remnant in the radio images.
Consequently,  it is difficult to obtain an accurate spectral index. However,
we estimate that between $\nu = 843$ MHz and $\nu = 4800$ MHz the remnant has a
rather flat spectrum with a spectral index $\alpha$ between $-$0.5 and 0
(defined as $S(\nu) \propto \nu^{\alpha}$ with $S (\nu)$ the flux density at
frequency $\nu$). The flatter radio-continuum spectrum is indicative of an
older remnant.

There is no infrared emission that can be clearly linked to the remnant. Bright
diffuse emission is seen towards the south of the remnant, indicating a
denser/dustier environment in that direction. We explore this further in
Sect.\,\ref{discussion_asymmetric}.

\label{results_morphology_snrd}
\paragraph{\snrd:} The source exhibits a rather atypical morphology in X-rays,
that can be described as ``triangular'' (Fig.\,\ref{fig_rgb_image_snrd}). It is
elongated along the NE--SW axis with a largest extent of $\sim 5.4$\arcmin\
(78.3~pc). The NE side of the triangle is brighter than the rest of the remnant
and extends $\sim 3.5$\arcmin\ (50.8~pc) along the SE--NW direction. This NE
``bar'' includes all the flux in the medium energy band, whilst the fainter SW
``tip'' appears softer. As nominal location of the remnant we took the incentre
of the triangle delineating the X-ray emission, which yields RA =
05\hour\,17\minute\,10.2\second, DEC = $-$67\degr\,59\minute\,03\second.

[\ion{S}{ii}] and H$\alpha$ lines are detected in the SW of the remnant,
closely following the ``tip'' of the X-ray emission, with strong \siiha\ ratios
(0.6--1.2, see Fig.\,\ref{fig_rgb_image_snrd} bottom left). The brighter and
harder X-ray ``bar'' lacks such optical emission. [\ion{O}{iii}]$\lambda$5007
\AA\ line emission is not observed anywhere in this remnant.

The presence of a point source close to the geometrical centre of \snrd\ is
evident in the image (Fig.\,\ref{fig_rgb_image_snrd}). We identified an
infrared/optical counterpart 2.4\arcsec\ away from the X-ray source, i.e. well
consistent with the typical position uncertainty of \xmm. The counterpart is
identified as SAGE J051710.30-675900.9 in the Spitzer catalogue of the LMC
\citep{2006AJ....132.2268M}. Based on its mid-IR colours, it was classified as
an active galactic nucleus (AGN) candidate by \citet{2009ApJ...701..508K}.
\citet{2013arXiv1305.6927K} later on spectroscopically confirmed the source as
a $z=0.427$ AGN. Therefore, we conclude that the central point source in \snrd\
is a background AGN rather than a compact stellar remnant. We discuss later the
morphology in greater detail (Sect.\,\ref{discussion_asymmetric}), in light of
the X-ray spectroscopy results (Sect.\,\ref{results_spectroscopy_snrd}).

The radio image of \snrd\ only shows weak, compact emission from the central
point source, consistent with the AGN classification discussed just above.
Bright diffuse 24 \textmu m emission is observed at the south-west of \snrd.
Infrared light intrinsically emitted by the remnant is however likely to be
masked by the emission of a nearby molecular cloud (as described in
Sect.\,\ref{discussion_asymmetric}). Besides this, two weak filaments outline
the eastern and western rims of the remnant. They are presented and discussed
in greater detail in Sect.\,\ref{discussion_asymmetric}.

\begin{figure*}[t]
    \begin{center}
    \includegraphics[angle=-90,width=0.533\hsize,viewport = 25 46 477 770,
clip]
    {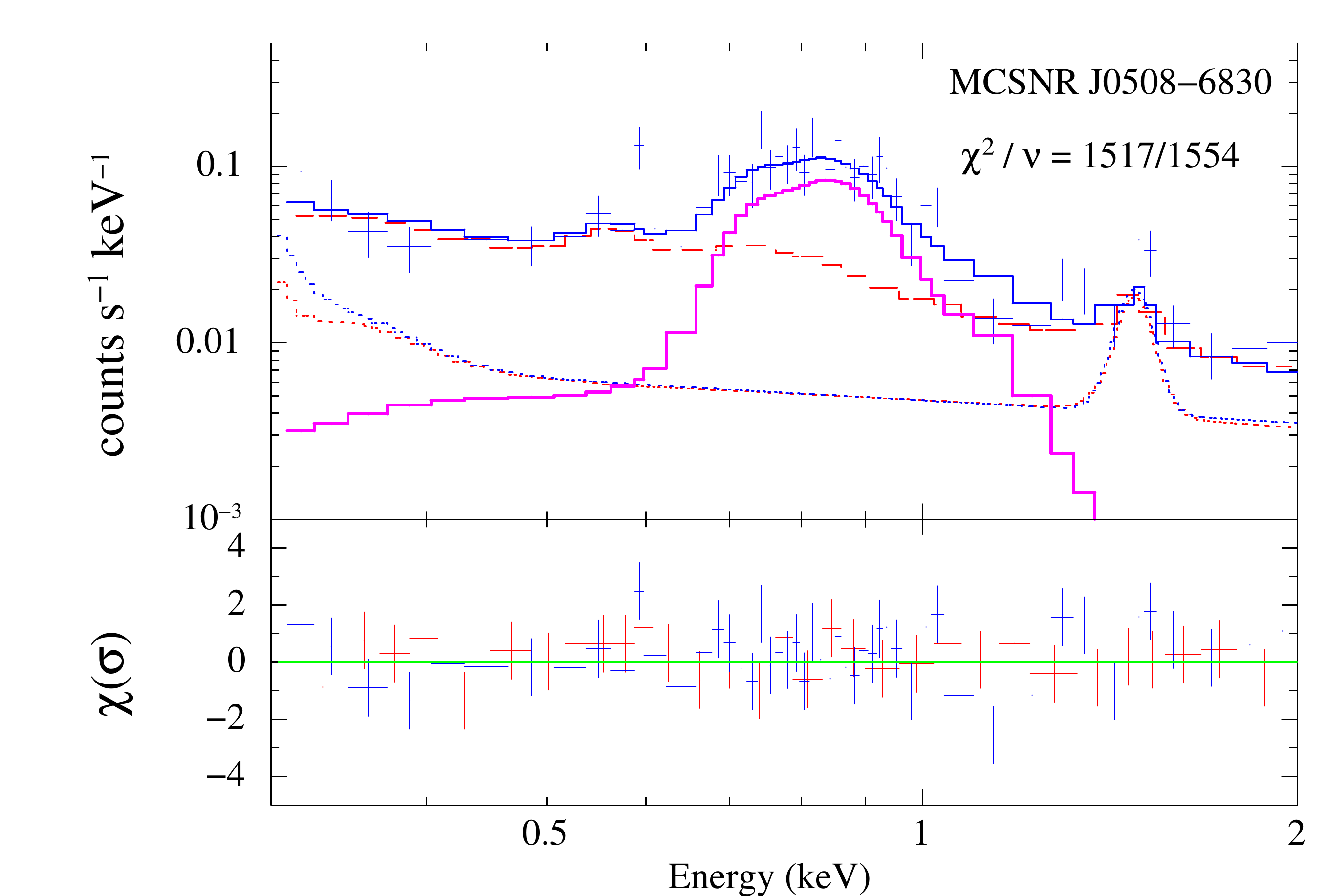}
    \includegraphics[angle=-90,width=0.457\hsize,viewport = 25 148 477 770,
clip]
    {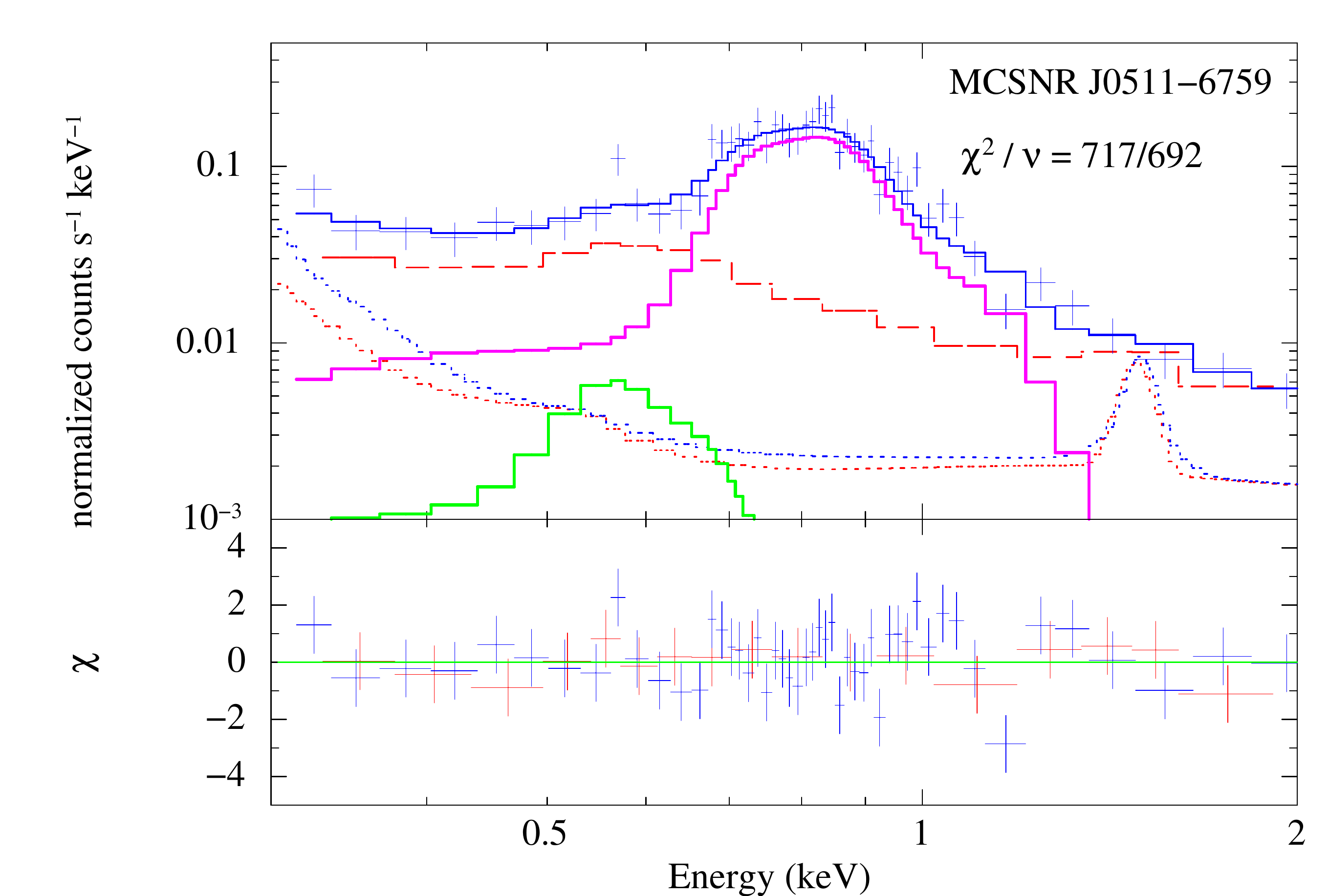}

    \includegraphics[angle=-90,width=0.533\hsize,viewport = 20 46 535 770,clip]
    {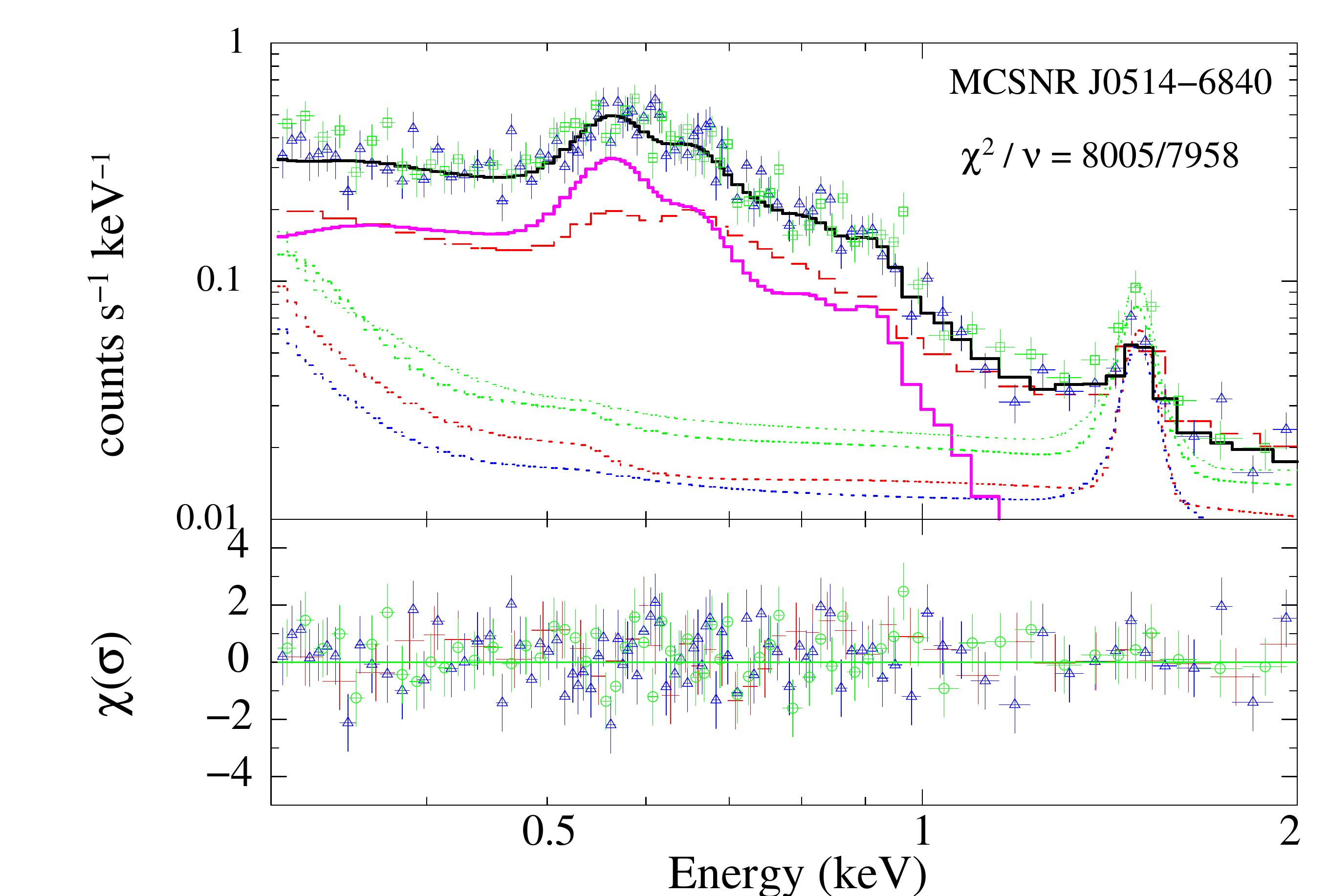}
    \includegraphics[angle=-90,width=0.457\hsize,viewport = 20 148 535 770,
clip]
    {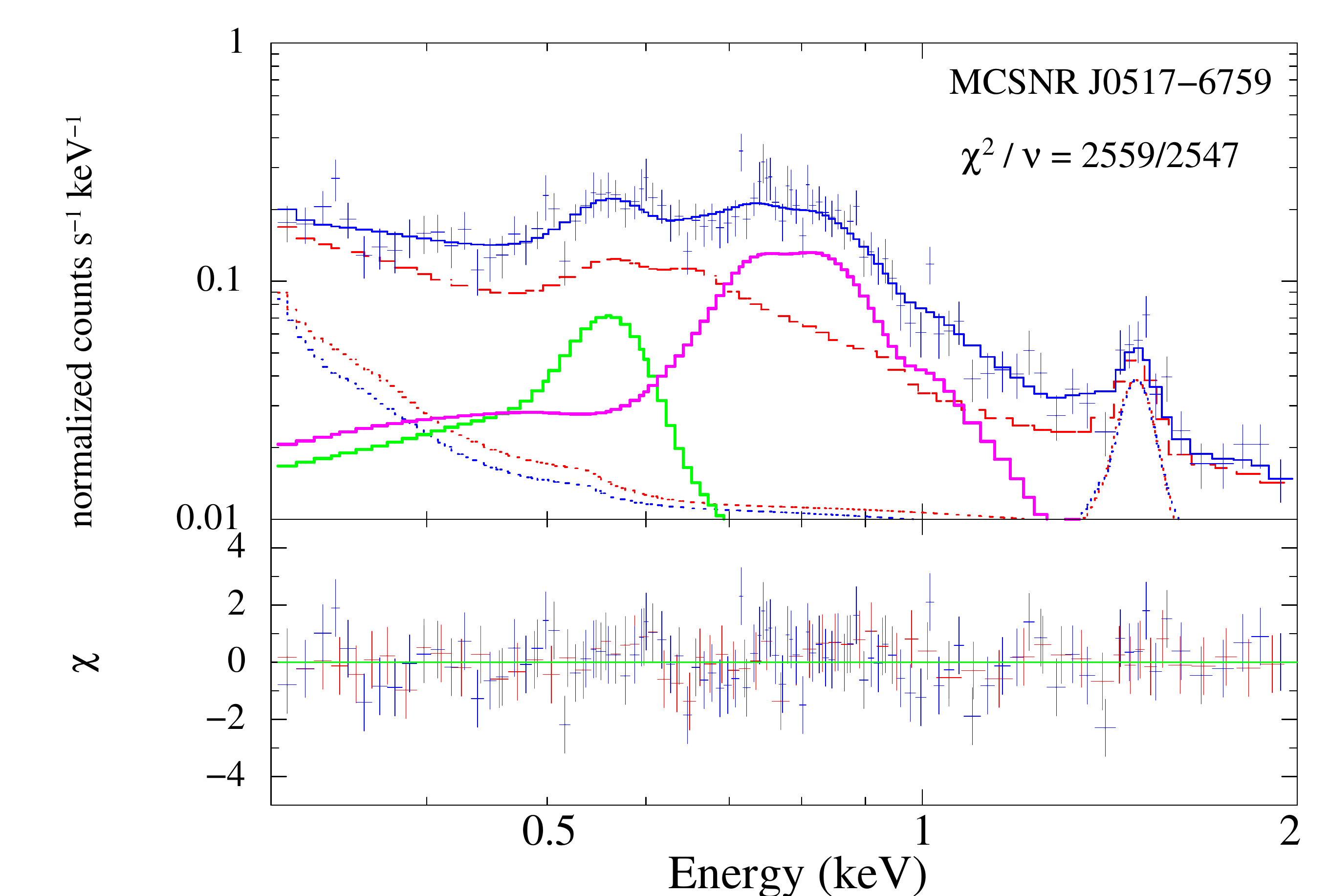}
    \end{center}
\sidecaption
\caption{X-ray spectra of the SNRs. Data extracted from the source region are
shown by blue data points, with the total (source+background) model as the
solid blue line. The red and blue dash-dotted lines show the instrumental
background model measured in the background and source extraction regions,
respectively. The X-ray+instrumental background model is shown by the dashed
red line. For clarity we do not show the data points from the background
extraction region but only the residuals of the fit (red points in the lower
panels). For \snrc, spectra from two overlapping observations are shown by the
green squares and blue triangles. The thick magenta lines show the source
emission component. When more than one component is in the SNR emission model
(see text for individual description), the secondary component is shown by the
thick green line. The residuals are shown in terms of $\sigma$.
}
    \label{fig_xray_spectrum}
\end{figure*}

\begin{table*}[t]
\caption{X-ray spectral results}
\label{table_spectral_results}
\centering
\begin{tabular}{@{}l @{\hspace{0.28cm}} @{\hspace{0.15cm}} c @{\hspace{0.15cm}}
@{\hspace{0.15cm}} c @{\hspace{0.15cm}} @{\hspace{0.15cm}} c @{\hspace{0.15cm}}
@{\hspace{0.15cm}} c @{\hspace{0.15cm}} @{\hspace{0.15cm}} c @{\hspace{0.15cm}}
@{\hspace{0.15cm}} c @{\hspace{0.15cm}} @{\hspace{0.15cm}} c @{\hspace{0.15cm}}
@{\hspace{0.15cm}} c}
\hline
\hline
\noalign{\smallskip}
Model & N$_{H\mathrm{\ LMC}}$ & $kT$ & $\tau$ & EM\tablefootmark{a} & 
O/H & Fe/H & $\chi ^2 $ / dof & $L_X$\tablefootmark{b}\\
&($10^{21}$ cm$^{-2}$) &(keV)&($10^{12}$ s\,cm$^{-3}$) &($10^{57}$ cm$^{-3}$)&
 & & & ($10^{34}$ erg\,s$^{-1}$) \\
\noalign{\smallskip}
\hline
\noalign{\smallskip}
\multicolumn{9}{c}{\snra}\\
\noalign{\smallskip}
\hline
\noalign{\smallskip}
\emph{vapec} (LMC abund.) & 0.06 ($<1.6$) & 0.66$_{-0.08}^{+0.06}$ & --- &
1.14$_{-0.2}^{+0.4}$ & 0.46 & 0.63 & 1508.92 / 1554 & 1.29 \\
\emph{vapec} (free abund.) & 0 & 0.71$_{-0.07}^{+0.06}$ &
--- &0.001$_{-0.0004}^{+0.7}$ & 0 & $>1.1$ \tablefootmark{c} & 1488.34 / 1554 &
0.90 \\
\emph{vpshock} (free abund.) & 0 & 0.60$_{-0.05} ^{+0.11}$ & 2.75 ($>0.47$) &
0.28$_{-0.24} ^{+1.8}$ & 0 & 4.12$_{-2.46} ^{+11.21}$ & 1519.42 / 1553 &
0.93 \\
\noalign{\smallskip}
\hline

\noalign{\smallskip}
\multicolumn{9}{c}{\snrb}\\
\noalign{\smallskip}
\hline
\noalign{\smallskip}
\emph{vapec} (LMC abund.) & 0 & 0.62$_{-0.04}^{+0.05}$ & --- & 1.9$\pm0.2$ &
0.46 &
0.63 & 732.85 / 693 & 2.16 \\
\emph{vapec} (free abund.) & 0 & 0.65$_{-0.04}^{+0.05}$ & --- &
0.12$_{-0.10}^{+0.8}$ &
0 & 11.4$_{-6.7} ^{+96.8}$ & 707.52 / 692 & 1.56 \\
\emph{vpshock} (free abund.) & 0 & 0.57$_{-0.05} ^{+0.06}$ & 0.75 ($>0.18$) &
0.60$_{-0.53} ^{+0.68}$ & 0 & 3.19$_{-1.78} ^{+3.41}$ & 700.96 / 691 &
1.74 \\
\noalign{\smallskip}
\hline

\noalign{\smallskip}
\multicolumn{9}{c}{\snrc}\\
\noalign{\smallskip}
\hline
\noalign{\smallskip}
\emph{vapec} (LMC abund.)  & 0 $(< 0.13)$ & 0.19$\pm0.01$ & --- & 10.2$\pm0.7$ &
0.46
& 0.63 & 8041.68 / 7960 & 3.42 \\
\emph{vapec} (free abund.) &0.6$_{-0.2} ^{+0.3}$&0.19$\pm0.01$ & --- &
17.8$_{-1.5}^{+1.6}$ & 0.30$_{-0.03} ^{+0.04}$ & 0.36$_{-0.20} ^{+0.26}$ &
7979.80 / 7958
& 3.81 \\
\emph{vpshock} (free abund.) & 0 ($<0.09$) & 0.30$\pm0.01$ & 0.26$_{-0.05}
^{+0.06}$ & 5.1$\pm0.3$ & 0.28$\pm0.03$ & 0.38$_{-0.11} ^{+0.13}$ &
8005.27 / 7957& 4.01\\
\emph{vsedov} (free abund.)  & 0 ($<0.07$) & 0.18 -- 0.25 \tablefootmark{d} &
--- ($>$) & 5.11 -- 11.2 & 0.36$\pm0.05$ & 0.34$\pm0.10$ & 7975.11/7958&
4.21 \\
\noalign{\smallskip}
\hline

\noalign{\smallskip}
\multicolumn{9}{c}{\snrd}\\
\noalign{\smallskip}
\hline
\noalign{\smallskip}
\multicolumn{9}{l}{2\emph{vapec} (LMC abund.)\,:}\\
\multicolumn{1}{r}{Cool component} & 3.5 & $< 0.14$ & --- & 112$_{-90}
^{+72}$  & \multirow{2}{*}{0.46 } & \multirow{2}{*}{ 0.63 } &
\multirow{2}{*}{2558.92 / 2547 } & 0.47 \\
\multicolumn{1}{r}{Hot component} & 0 & 0.59$_{-0.04} ^{+0.05}$ & --- &
1.85$_{-0.20} ^{+0.16}$ & & & & 1.96 \\
\noalign{\smallskip}
\hline

\end{tabular}
\tablefoot{Best-fit parameters of the various source models (details are in
Sect.\,\ref{results_spectroscopy}). Uncertainties are given at the 90\,\%
confidence level. Parameters with no uncertainties were frozen. The abundances
of O and Fe are given relative to the solar values as listed in
\citet{2000ApJ...542..914W}. The $\chi^2$ and associated degrees of freedom
(dof) are also listed.
\tablefoottext{a}{Emission measure $\int n_e n_H dV$.}
\tablefoottext{b}{Absorbed luminosity in the 0.3--5 keV band.}
\tablefoottext{c}{$3\sigma$ lower limit.}
\tablefoottext{d}{See Sect.\,\ref{results_spectroscopy_snrc} for a description
of the uncertainties of the \emph{vsedov} model.}
}
\end{table*}

\subsection{X-ray spectroscopy}
\label{results_spectroscopy}

\label{results_spectroscopy_snra}
\paragraph{\snra:} 
The X-ray spectrum of this remnant is shown in Fig.\,\ref{fig_xray_spectrum},
top left. The most striking feature is the large Fe L-shell bump which dominates
the X-ray emission.

Despite the faintness of the source, we could obtain meaningful best-fit
parameters and uncertainty ranges with the simple spectral models. The results
are listed in Table\,\ref{table_spectral_results}. The best fits were obtained
for temperatures of about 0.6--0.7 keV and a low absorption, consistent with
$N_H \sim 0$ cm$^{-2}$ (90\,\% C.L. upper limit of $1.8 \times 10^{21}$
cm$^{-2}$). We note that for low $N_H$ values ($\lesssim 10^{21}$ cm$^{-2}$),
absorption effects are small and mostly affect photons below 0.5  keV. Since the
source shows no significant emission below this energy, $N_H$ cannot be
efficiently constrained. We fixed the LMC $N_H$ to 0 cm$^{-2}$ for the rest of
the analysis, stressing that this does not influence the results we present
below. No significant effects of non-equilibrium ionisation were detected\,: the
ionisation age was close to the upper limit available in the \emph{vpshock}
model.

The fits greatly improved when the Fe abundance was let free, and improved
marginally if O abundance was free as well. The O abundance tended towards 0,
but was essentially unconstrained (upper limit of $\sim$20 times the solar
value). This happens because at the best-fit temperature ($\sim 0.6$ keV), which
is set by the shape of the iron L-shell bump, the oxygen emissivity is
relatively low. We therefore cannot well constrain this parameter. The Fe
abundance was found to be greatly in excess of the average LMC value, or even
solar value. The upper limit of Fe/Fe$_{\sun}$ is very high or unconstrained
because of the degeneracy between this parameter and the normalisation of the
\emph{vapec} (or \emph{vpshock}) component. Indeed, since iron is almost
the
only contributor to the spectrum, the fitting procedure cannot distinguish
between a higher iron abundance and lower emission measure, or
\emph{vice-versa}. The low contribution of oxygen to the spectrum and
predominance of iron is investigated further in a multi-component plasma
analysis (Sect.\,\ref{results_spectroscopy_multicomp}).

\label{results_spectroscopy_snrb}
\paragraph{\snrb:}
The X-ray spectrum of this remnant (Fig.\,\ref{fig_xray_spectrum}, top right)
resembles that of \snra, as expected from their morphological and X-ray colour
similarities. It is also dominated by an Fe L-shell bump and has an even lower
contribution from oxygen lines.

The best-fit \emph{vapec} and \emph{vpshock} models are obtained for plasma
temperatures of 0.64 keV and 0.56 keV, respectively (see
Table\,\ref{table_spectral_results}). Initial trial fits were made with a free
LMC absorption column. They consistently returned 0 cm$^{-2}$ as the best-fit
value for $N_H$, although the 90\,\% C.L. upper limit of $2.1 \times 10^{21}$
cm$^{-2}$ for $N_H$ is quite significant. In the rest of the analysis we fixed
$N_H$ to 0 cm$^{-2}$ (see the caveat on $N_H$ presented above for \snra).

Again, no oxygen was formally required, whilst an Fe abundance greatly in excess
of the solar value was needed. For the same reason as for \snra\ we investigated
the iron-rich nature of the sourve with a multi-component plasma analysis
(Sect.\,\ref{results_spectroscopy_multicomp}).

The ionisation age $\tau$ in the plane-parallel shock model was high (best-fit
value of $8.7 \times 10^{11}$ s\,cm$^{-3}$). Its high 90\,\% C.L. lower limit
($1.6 \times 10^{11}$ s\,cm$^{-3}$) and its unconstrained upper limit suggests
that the X-ray emitting plasma in \snrb\ is close to or at collisional
ionisation equilibrium.

\label{results_spectroscopy_snrc}
\paragraph{\snrc:} Spectra from the two observations of the source were fit
simultaneously. The parameters of the SNR component in both spectra were tied
together. The astrophysical background components also shared the same
parameters, allowing only for a constant factor between the two sets of spectra.
Only the (detector position-dependent) instrumental background and
(time-dependent) SPC components had different parameters for each observation.

Good fits were obtained for relatively soft temperatures of 0.2--0.4 keV,
depending on the model used (see Table\,\ref{table_spectral_results}). When O
and Fe abundances were let free to vary, the fits improved significantly (e.g.
$\chi ^2 / \nu = 7979.80 / 7958 $ instead of $\chi ^2 / \nu = 8041.68 / 7960$).
However, the best-fit values for O/O$_{\sun}$ and Fe/Fe$_{\sun}$ were both only
reduced by a factor of $\sim 0.6$ compared to the value given in
\citet{1992ApJ...384..508R}, being rather consistent with those from
\citet{1998ApJ...505..732H}, whilst the ratio O/Fe remained well within the
uncertainties of the LMC ISM value given in the two latter references. This
indicates that the SN ejecta have no significant contribution to the X-ray
spectrum, which is dominated by the swept-up ISM. This justifies \emph{a
posteriori} that the remnant is indeed well in the Sedov phase. It also means
that no typing of the SN progenitor can be achieved through the spectral
analysis.

We show the X-ray spectrum fitted with the Sedov model in
Fig.\,\ref{fig_xray_spectrum} (bottom left). The formal best-fit parameters
with this model are listed in Table\,\ref{table_spectral_results}. The best-fit
temperature is rather low and the ionisation age rather high ($\sim 2 \times
10^{12}$ s\,cm$^{-3}$). These two parameters are however relatively poorly
constrained. We investigated the $kT$ vs. $\tau$ parameter space: equally
acceptable fits are allowed both for low temperatures ($\sim 0.2$ keV) with high
ionisation ages ($\sim 2 \times 10^{12}$ s\,cm$^{-3}$), and for higher
temperatures ($\sim 0.25 - 0.3$ keV) with lower ionisation ages (a few $10^{11}$
s\,cm$^{-3}$). This degeneracy is explained to some extent to the statistics we
have: only lines from a limited set of elements (O, Fe, and possibly Ne) are
detected and can be used to constrain those parameters. Another contributor is
possibly that the assumption of the Sedov model of a uniform ambient medium does
not hold, resulting in asymmetric evolution and varying plasma conditions. The
presence of such an inhomogeneous ISM is supported by the optical image, as only
the southern edge of the remnant emits lines; as for the X-ray image, the
remnant is (marginally) brighter in the southern half.

Guided by the morphology of \snrc\ (Sect.\,\ref{results_morphology_snrc},
Fig.\,\ref{fig_rgb_image_snrc}), we extracted spectra from the southern and
northern halves, in order to look for possible plasma properties or column
density (or both) variations across the remnant. In all this analysis, we
assumed
the ISM to have a homogeneous chemical composition and abundances were fixed at
their best-fit values. We divided the SNR along the ``dark lane'' that crosses
the remnant's equator. First using CIE models, it was possible to constrain the
temperature and $N_H$ of both spectra, despite the degraded statistics. We found
that they had similar temperature (0.18--0.22 keV), but that $N_H$ was
significantly higher in the south than in the north ($\sim 1.3 \times 10^{21}$
cm$^{-2}$ vs. $\sim 0.3 \times 10^{21}$ cm$^{-2}$). Using either the
\emph{vpshock} or Sedov model, we found (roughly five times) higher ionisation
ages in the south spectrum as compared to the north, which is again an
indication of an inhomogeneous ISM. More specifically, it indicates a density
gradient increasing southwards. We interprete these results as environmental
effects in Sect.\,\ref{discussion_asymmetric}.

If we assume the Sedov self-similar solution for \snrc, we can calculate its
properties. Given the issues discussed in the above paragraphs, there are
concerns that this model, which assumes a spherical symmetry and homogeneous
ISM, might yield incorrect results. However, using the best-fit parameters from
the integrated spectrum as a measure of the properties averaged over the
remnant, we can still obtain rough but useful estimates of important numbers
(e.g. age, density, etc...). Alternatively, we can compute the physical
properties of the remnant with parameters derived when fitting the north and
south spectra, and use these as limiting cases.

The normalisation of the \emph{vsedov} model is proportional to the volume
emission measure $\int n_e n_H dV$, which can be rewritten as a function of the
pre-shock ambient hydrogen density $n_{H,0}$\,:
\begin{equation}
    {\rm EM} = \left( \frac{n_e}{n_H} \right) n_{H,0}^2 \frac{4\pi }{3} R_{S}^3
\int _0 ^1 3 \left( \frac{\rho (r)}{\rho_0} \right)^2 r^2 dr
\end{equation}
where $R_S$ is the shock radius and $r$ the normalised radius ($R/R_S$). To
evaluate the integral one can use the approximation of
\citet{1975ICRC...11.3566K} for the normalised mass distribution in the Sedov
model (his equation 7.19); numerical integration then gives $\approx 2.07$.
Given the radius $R_S$ of 26.5 pc and $n_e / n_H \approx 1.2$ (for a fully
ionised, 0.5 $Z{\sun}$ plasma), we obtain pre-shock densities $n_{H,0} = (0.03
- 0.05)$ cm$^{-3}$, using the integrated spectrum. We can then estimate the
mass swept-up by the SNR shock as $M = (4\pi /3) R_S^{3} 1.4 m_p n_0$, with
$n_0$ the ambient pre-shock ion density ($\approx 1.1 n_{H,0}$) and $m_p$ the
proton mass. $M$ is in the range (90 -- 150) \msun.

To estimate the dynamical age of the remnant $t_{dyn}$, we assume strong shock
conditions and full equilibration between electron and ions. Then, the shock
velocity $v_S$ is related to the X-ray temperature $T_X$ by
\begin{equation}
    v_S = 
    \sqrt{ \frac{16 \ k T_X}{3\  \mu \ m_p} } 
    \approx
    914 \left( \frac{k T_X}{\mathrm{1 \ keV}} \right) ^{1/2} \mathrm{km\,
s}^{-1}    ,
\end{equation}
where $k$ is the Boltzmann constant and $\mu m_p$ is the mean molecular
mass ($\approx 0.61 m_p$). The shock velocity is therefore in the range (390 --
470) km\,s$^{-1}$. This is then related to $t_{dyn}$, the dynamical age of the
remnant from Sedov's similarity relation $v_S = 2 R_S / 5 t_{dyn}$. We found
$t_{dyn}$ to be 22 -- 27 kyr.  The flatter radio-continuum spectrum is
consistent with this quite advanced age. The explosion energy is given by $E_0 =
1.4 m_p n_0 R_S^5 / 2.02 t_{dyn}^2$, yielding $E_0 = (0.2 - 0.5) \times 10^{51}$
erg.

Finally, $t_{ion}$, the age of the remnant derived from the ionisation age
$\tau$ was about an order of magnitude larger than the dynamical age. The value
of $\tau$ is however relatively poorly or even not constrained. We note that
this can happen if elements contributing to the thermal spectrum are close to
CIE. Indeed, as shown by \citet{2010ApJ...718..583S}, at $kT \sim 0.2$ keV all
astrophysically abundant elements reach ionisation stages close to their
equilibrium values on timescales of the order of $10^{11}$~s\,cm$^{-3}$ to
$10^{12}$~s\,cm$^{-3}$. The non-equilibrium effects in the plasma are therefore
small and the ages derived highly uncertain. Values as low as
$2 \times 10^{11}$~s\,cm$^{-3}$, which give $t_{ion} \sim t_{dyn}$, are still
acceptable within the $3\sigma$ uncertainties.

\label{results_spectroscopy_snrd}
\paragraph{\snrd:}
We started by analysing the integrated spectrum of \snrd, excluding only the
central background AGN. An initial fit with a one-temperature CIE model
(\textit{vapec}) with LMC abundances failed to reproduce the spectrum, as
indicated by strong residuals. Namely, the ``best-fit'' model, with $kT \sim
0.5$ keV, could reproduce the Fe L-shell emission (between 0.7 and 1.1 keV) but
underpredicted the data around 0.5--0.6 keV (dominated by K lines of
\ion{O}{VII}) whilst predicting too much flux at 0.6--0.7 keV (dominated by the
\ion{O}{VIII} Lyman series). In other words, the temperature constrained by
the Fe emission is too high for oxygen. This issue could neither be resolved by
using an NEI model, nor by changing the O/Fe abundance balance, because at $kT
\sim 0.5$ keV oxygen is mostly in the H-like ionisation stage
\citep{1982ApJS...48...95S}, and simply increasing the O abundance would
overproduce $\sim$0.65 keV emission even more.

Driven by this result and by the morphological analysis of the source
(Sect.\,\ref{results_morphology_snrd}), we concluded that a two-temperature
model was required. We used two \textit{vapec} models with distinct temperatures
and absorption columns, but both with LMC abundances. This time we obtained
satisfactory fits, with no systematic residuals. The integrated spectrum fitted
with this model is shown in Fig.\,\ref{fig_xray_spectrum} (bottom right).
Best-fits were obtained with a ``hot'' ($kT_{\rm{hot}}\sim 0.6$ keV) and
``cool'' ($kT_{\rm{cool}}\sim 0.1$ keV) component. The hot component models the
Fe and \ion{O}{VIII} emission, whilst the low-temperature component accounts for
the extra \ion{O}{VII} emission. Although the absorption was poorly determined,
the ``cool'' component required a significantly higher $N_H$ (0.6--8.3$\times
10^{21}$~cm$^{-2}$) than the ``hot'' one, which had a best fit-value of $\sim
0$ cm$^{-2}$ and an upper limit of 1.7$\times 10^{21}$~cm$^{-2}$. We chose to
fix the absorption for the ``hot'' component to 0 cm$^{-2}$, whilst for the
low-temperature component we fixed the $N_H$ to 3.5$\times 10^{21}$~cm$^{-2}$,
as measured from the \ion{H}{I} map of \citet{2003ApJS..148..473K}. We give the
best-fit parameters of this model and the luminosity of both components in
Table\,\ref{table_spectral_results}.

We then proceeded to apply this model to spectra extracted from various regions
of the SNR., namely from the NE ``bar'' and the SW ``tip''
(Fig.\,\ref{fig_rgb_image_snrd}). Only normalisations and temperatures of the
two components were allowed to change. The best-fit temperatures from the NE
and SW spectra were the same as in the integrated spectrum. As expected from
the images, we found that $\sim$ 80\% of the flux of the ``hot'' component is
from the NE ``bar'', and $\gtrsim$ 90\% of the ``cool'' emission is in the SW
tip. Scenarios for the origin of this peculiar morphological and spectral
features are presented in Sect.\,\ref{discussion_asymmetric}.

\subsection{Multi-component plasma fits}
\label{results_spectroscopy_multicomp}
The X-ray emission of \snra\ and \snrb\ is dominated by iron, with a possible
minimal contribution from oxygen. To investigate this further, we modeled these
sources with a multi-component plasma, each component representing emission
from a single element. This approach has been used in the past and allows,
under some assumptions, to calculate the mass of the supernova nucleosynthesis
products \citep[e.g.][]{2003ApJ...582L..95H,2010A&A...519A..11K,bozzetto2013}.

As initial spectral fits showed (Sect.\,\ref{results_spectroscopy}), the
interior plasma is likely to be in CIE. Consequently, we used \emph{vapec}
models. The abundance of each element in its respective component was set to
$10^9$ the solar values, thus making sure we approximate a pure-metal component.
Only plasma composed of Fe and O needed to be included in the fit. The
temperature of the oxygen plasma ($kT_{\mathrm{O}}$) was not well constrained,
given the very small contribution of this element. Therefore, we tied the
temperature of this component to that of the Fe component. This is expected if
these two elements are co-spatial. We also tried fits with $kT_{\mathrm{O}}$
fixed at the peak emissivity temperature of the strongest oxygen lines in the
0.3--1 keV range (i.e. $kT_{\mathrm{O}} = 0.17$ keV). This turned out to have
very little influence on the results, which we give for the two remnants in
Table\,\ref{table_spectral_results_multicomp}. The fits returned a zero
normalisation of the oxygen component in both \snra\ and \snrc, showing the
minimal contribution of O in the emission of the two remnants, as expected from
the spectral analyses described above.

\begin{table}[t]
\caption{Results of the multi-component plasma fits for \snra\ and \snrb.
The derived mass of iron is given for two level of H admixture in the ejecta, as
described in Sect.\,\ref{results_spectroscopy_multicomp}.}
\begin{center}
\label{table_spectral_results_multicomp}
\begin{tabular}{l c r}
\hline
\hline
\noalign{\smallskip}
  \multirow{2}{*}{Fit parameter} & J0508$-$6830 & J0511$-$6759 \\
& \multicolumn{2}{c}{Value} \\
\noalign{\smallskip}
\hline
\noalign{\smallskip}
$kT_{\rm{Fe, O}}$ (keV) & $0.71\pm0.06$ & 0.65$_{-0.03} ^{+0.06}$ \\
EM$_{\rm{Fe}} \times n_{\rm{Fe}}/n_H$ ($10^{57}$ cm$^{-3}$) & 0.8$\pm 0.1$ &
1.5$\pm0.1$ \\
EM$_{\rm{O}} \times n_{\rm{O}}/n_H$ ($10^{57}$ cm$^{-3}$) & 0 $(<0.7)$ &
0 $(<0.8)$\\
$\chi^{2}/\nu$& 1516.99/1554 & 717.46/692 \\
\noalign{\smallskip}
$V$ ($10^{59}$ cm$^{3}$) & 4 & 3 \\
$M_{\rm{Fe}}$ (\msun, case I) & 1.03$\pm0.08$ & 1.19$\pm0.05$ \\
$M_{\rm{Fe}}$ (\msun, case II) & 0.51$\pm0.04$ & 0.59$\pm0.03$  \\
\noalign{\smallskip}
\hline
\end{tabular}
\end{center}
\end{table}

The normalisation of each component is proportional to the emission measure
EM$_X$ (given in terms of $n_e n_H V$ for each component $X$). Therefore, given
a knowledge of the number ratios $n_X/n_H$ and $n_e/n_X$ for an element $X$, we
can derive the mass $M_X$ of that element produced by the supernova using
\begin{equation}
    M_X = \sqrt{\frac{V_X \mathrm{EM} \ (n_X/n_H)}{(n_e/n_X)}} m_\mathrm{U} A_X
\end{equation}
\citep[e.g.][]{2010A&A...519A..11K,bozzetto2013}, where $m_\mathrm{U}$ is the
atomic mass unit. $A_X$ is the atomic mass of element $X$ and $V_X$ the volume
it occupies. For \snrb\ we assumed a spherical morphology with a radius of 13.5
pc (Sect.\,\ref{results_morphology}), and therefore $V_X = 3 \times 10 ^{59}$
cm$^{3}$. The volume of \snra\ is derived assuming an ellipsoidal morphology,
with semi-major and minor axes of 16.7 pc and 13.1 pc. As a third semi-axis we
took the average of the two others (i.e. 14.9 pc), yielding $V_X = 4 \times 10
^{59}$ cm$^{3}$. This volume would be 13\% higher or lower in the case of an
oblate or prolate morphology, respectively.

\begin{figure*}[t]
    \begin{center}
    \includegraphics[width=0.49\hsize]
    {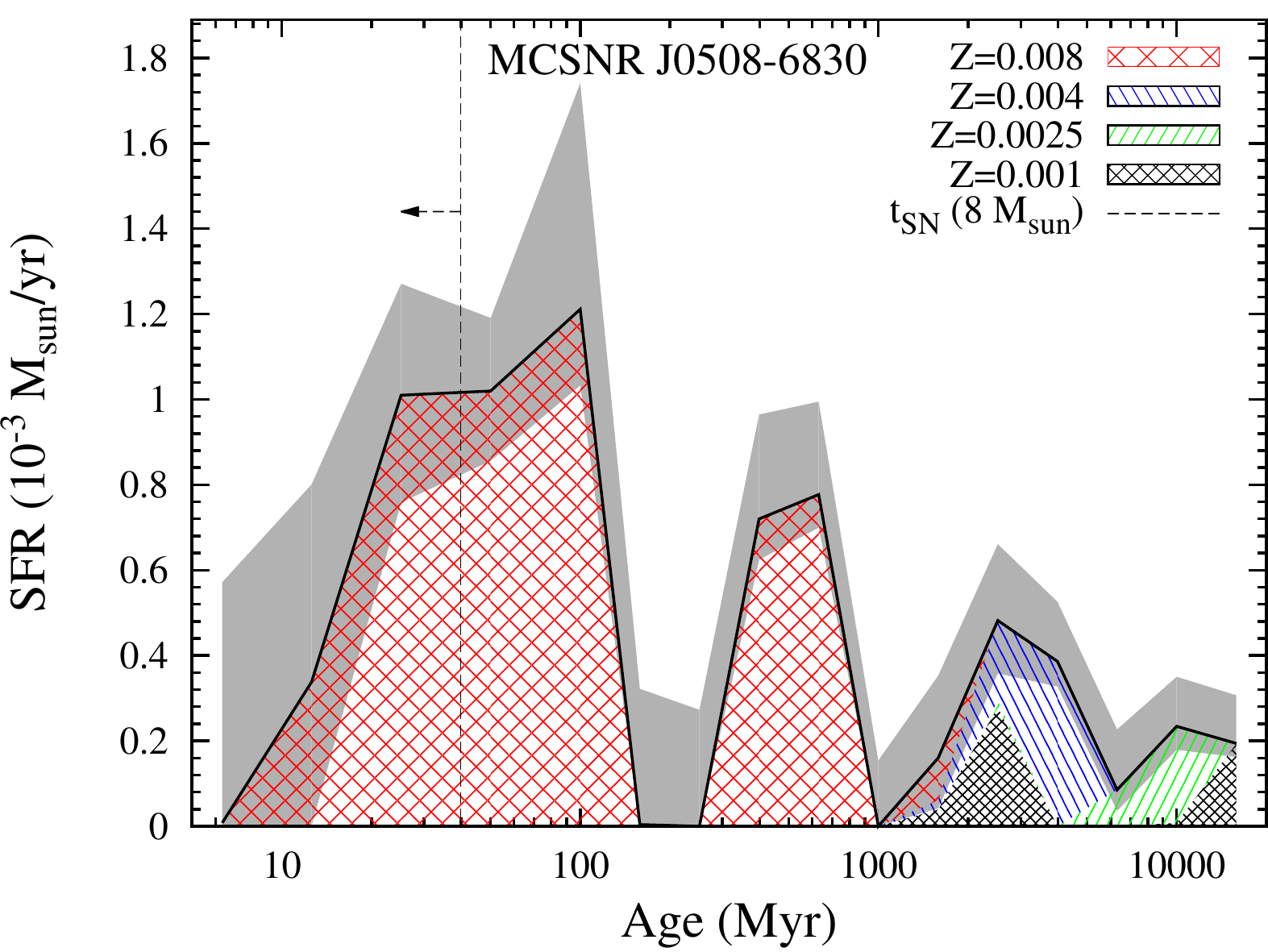}
    \includegraphics[width=0.49\hsize]
    {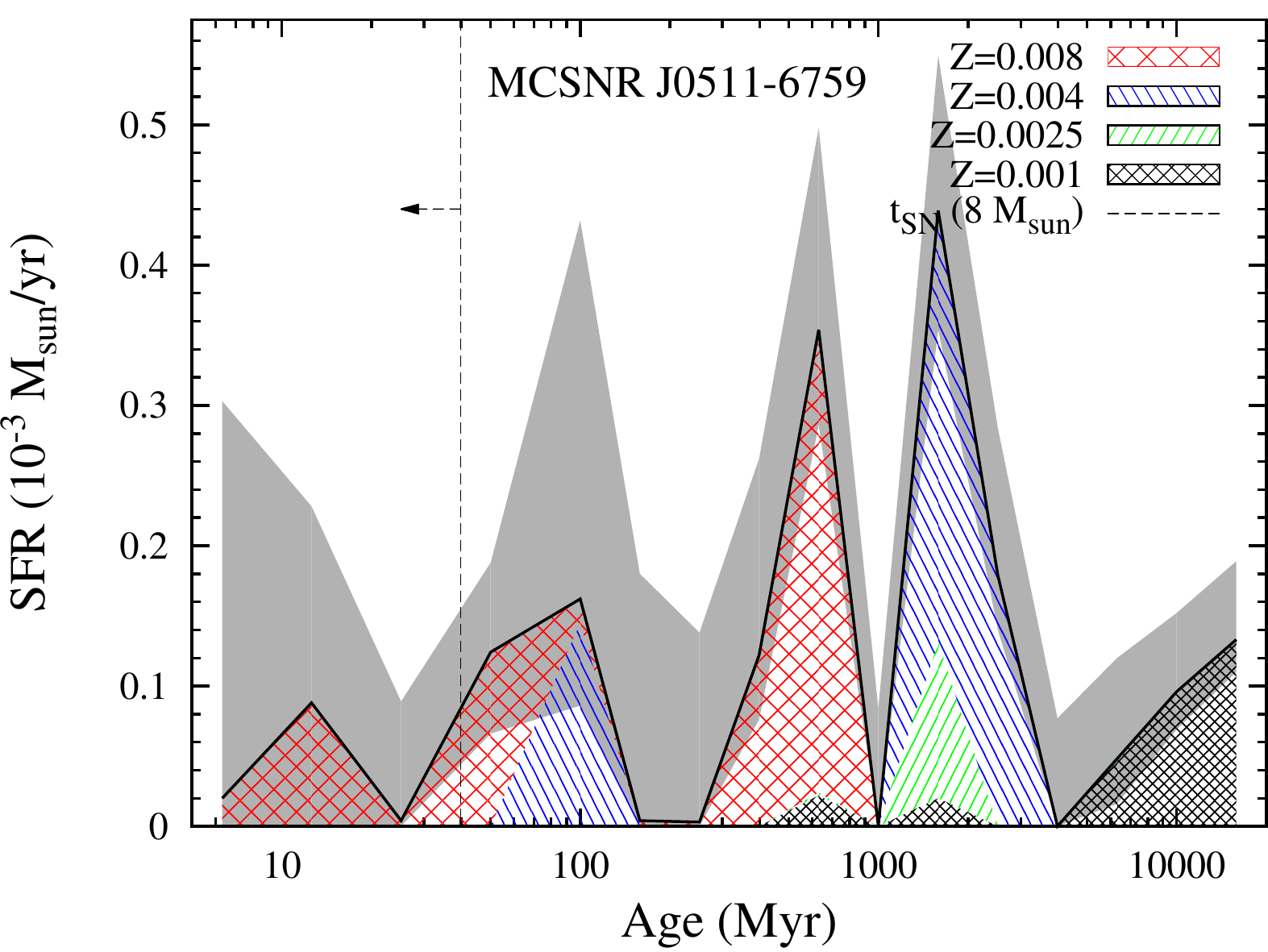}

    \includegraphics[width=0.49\hsize]
    {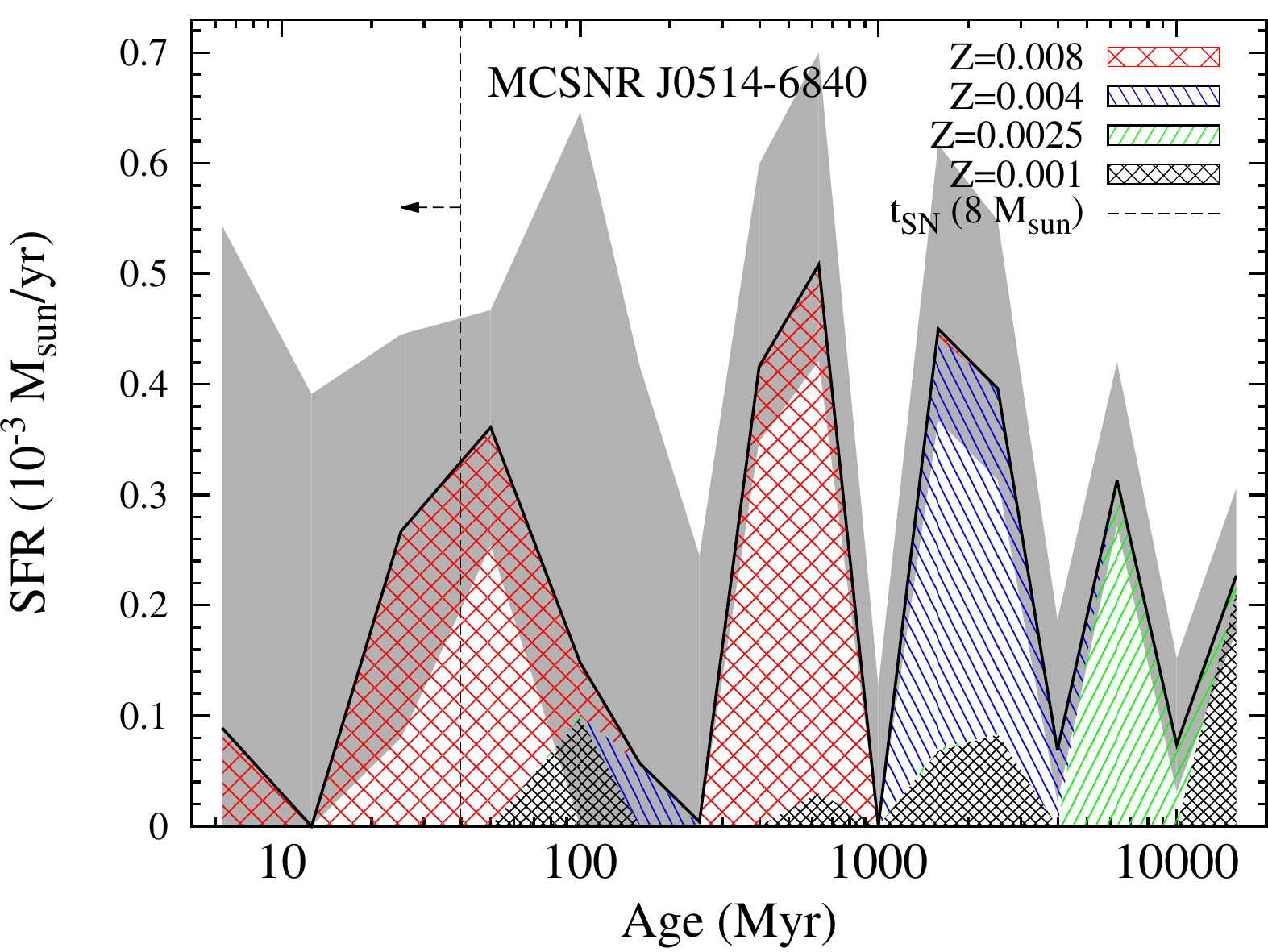}
    \includegraphics[width=0.49\hsize]
    {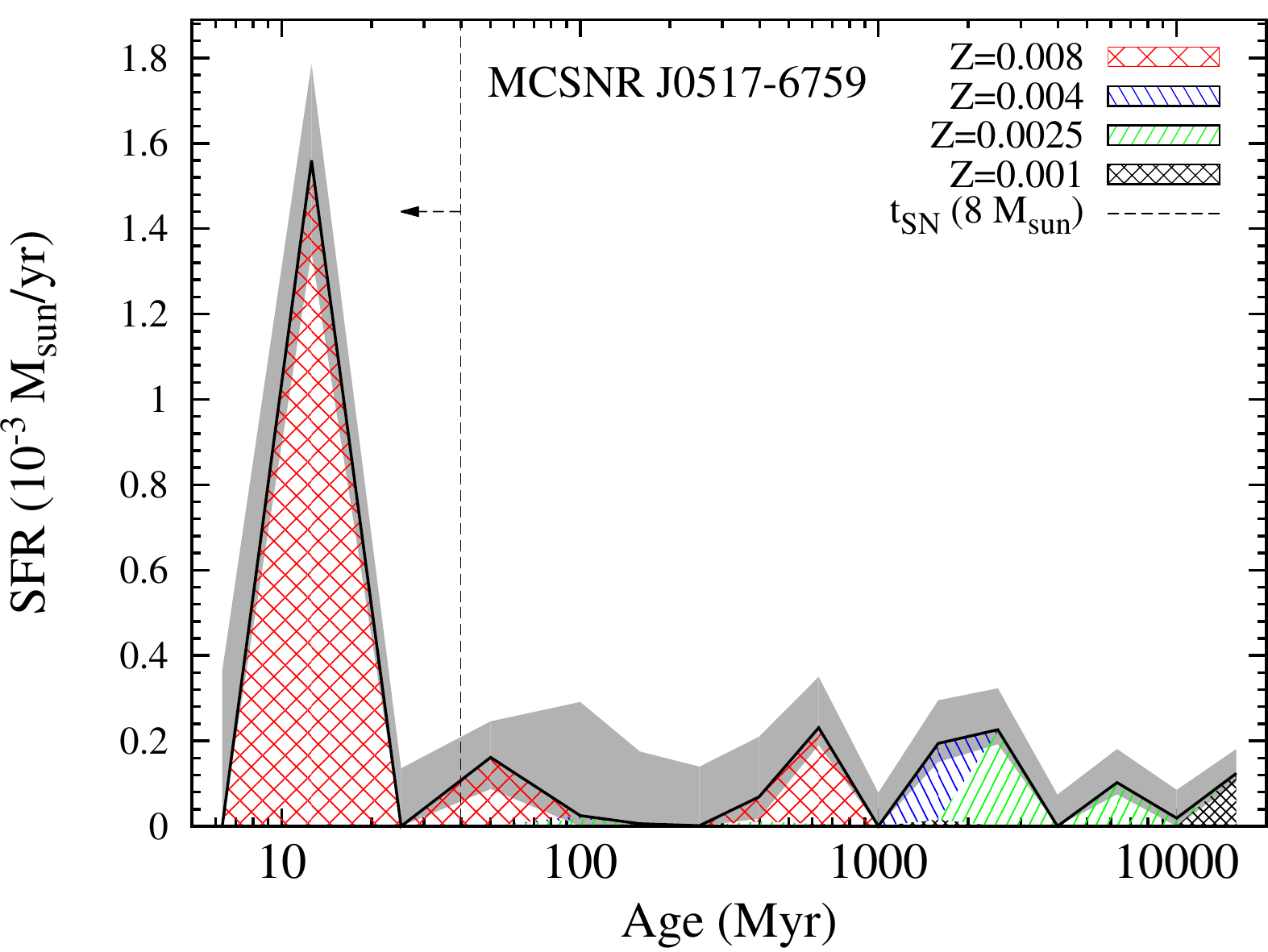}
    \end{center}
\caption{Star formation history around the remnants. Data are taken from
\citet{2009AJ....138.1243H}. The star formation rate in four metallicity bins
are plotted against lookback time. The errors (combining all metallicities) are
shown by the gray shading. The vertical dashed line at 40 Myr indicates the
maximal lifetime of a CC SN progenitor. Note the changing vertical scale.}
    \label{fig_sfh}
\end{figure*}

The main uncertainty for estimating $M_X$ is in the ratio of electron-to-ion
$n_e/n_X$. We follow the prescription of \citet{2003ApJ...582L..95H} and
consider two limiting cases. In the first one (case I) the emission originates
purely from ejecta, without admixture from hydrogen. Considering that Fe
dominates the ejecta, with only minimal contribution from oxygen to the pool of
electrons, this means $n_e/n_{\rm{Fe}}$ is only set by the average ionisation
state of iron. At this temperature we take $n_e/n_{\rm{Fe}} = 18.3$
\citep{1982ApJS...48...95S}. The second physically plausible case (case II)
assumes that a similar mass of H has been mixed with the iron ejecta. Therefore,
there are 56 H atoms (contributing each with one electron) per Fe atom, and
$n_e/n_{\rm{Fe}}$ is 74.3. We give the best-fit parameters and the derived iron
mass in both cases in Table\,\ref{table_spectral_results_multicomp}.

\subsection{Local stellar population}
\label{results_sfh}
We now describe the SFH in the neighbourhood of the four SNRs. We use the
spatially resolved SFH map of \citet{2009AJ....138.1243H}. The map was derived
from the $UBVI$ photometric survey of \citet{2004AJ....128.1606Z} through
colour-magnitude diagram fitting. The results are given in terms of star
formation rate (SFR, in \msun\, yr$^{-1}$) in 13 time bins and four metallicity
bins, for 1380 cells, most of them having a size of 12\arcmin $\times$
12\arcmin.

Because stars might drift away from their birth place, one potentially important
caveat is that the SFH of a cell hosting a SNR may be derived from stars having
no physical connection with the SNR progenitor. For a detailed discussion on the
relevance of local stellar populations to the study of progenitors, we point to
\citet{2009ApJ...700..727B}. However, we stress that most of the information we
can gain from the study of the local SFHs, in the context of broadly typing a
remnant, is contained in the most recent time bins. Namely, the presence of
recent star formation is a strong necessary (but not sufficient, see the
introduction) condition to tentatively type a remnant as having a CC origin.
Conversely, the lack of recent star-forming activity favours a thermonuclear
origin (see Sect.\,\ref{discussion_typing}). 

For each SNR we plot the SFR of the cell including the remnant
(Fig.\,\ref{fig_sfh}, as a function of lookback time and metallicity. \snra\ and
\snrc\ are located in the region coined ``outer bar'' by
\citet{2009AJ....138.1243H}, and they share some properties: the SFR is moderate
in the recent past and is declining at times $t <$ 50 Myr. The quiescence
periods at 250 Myr and 1 Gyr are rather deep. The SFR around \snrc\ has its
stronger peaks at $\sim$ 500 Myr and 1.6 Gyr, whilst \snra\ exhibits an even
stronger peak at 100 Myr.

\snrb\ and \snrd\ reside in the ``Northwest void'' region of
\citet{2009AJ....138.1243H}, characterised by a low SFR throughout all epochs
and no significant enhanced star formation activity period (note the different
vertical axis scale of the SFH plots). However, the cell including \snrd\ shows
a strong peak at 12 Myr. We checked the neighbouring cells to find that this
recent SFR enhancement is concentrated around the remnant rather than being a
larger scale feature. \snrd\ is the only remnant of the sample whose SFH is
dominated by a burst of recent star formation.

\section{Discussion}
\label{discussion}

With these four new objects, the SNR population of the LMC now amounts to more
than 60 (54 in \citealt{2010MNRAS.407.1301B}, plus
\citealt{2012A&A...539A..15G}; \citealt{2012A&A...546A.109M};
\citealt{2013A&A...549A..99K}; \citealt{2013MNRAS.432.2177B,bozzetto2013}). The
SNRs have sizes in the upper range of the distribution of diameter of LMC SNRs
\citep{2010MNRAS.407.1301B}, as expected from their fairly advanced
evolutionary stage.

The non-detection in radio of three out of four objects (namely, \snra, \snrb,
and \snrd) is rare for LMC SNRs, as all known remnants either show
radio-continuum emission \citep[e.g.][]{2010MNRAS.407.1301B} or available
observations are impacted by bright objects in the remnants' neighbourhood.
This lack of detection would imply old remnants, with ages in excess of 20--30
kyr. In these cases the SNRs become very faint, and coupled with the likely low
density environment in which they reside, it is not surprising that we are
unable to detect them at any radio frequencies. Future radio-continuum
observations in this lower frequency range, potentially with the Murchison
Widefield Array \citep[MWA, ][]{2013PASA...30....7T}, may be able to detect
such
faint emission.

We now discuss the most interesting features of each
remnant, namely\,:
for \snrc\ and \snrd, the observed effect of an inhomogeneous ISM; for all four
SNRs, the type of SN origin; for \snra\ and \snrb, the iron-rich nature.

We now discuss the most interesting features of each
remnant, namely\,:
for \snrc\ and \snrd, the observed effect of an inhomogeneous ISM; for all four
SNRs, the type of SN origin; for \snra\ and \snrb, the iron-rich nature.

\subsection{Environmental effects and asymmetric evolution}
\label{discussion_asymmetric}
\snrc\ displays a subtle variation of X-ray colour along its north-south axis
(Fig.\,\ref{fig_rgb_image_snrc}), being harder towards the south. The X-ray
spectral analysis in Sect.\,\ref{results_spectroscopy_snrc} strongly suggests
that this is a foreground extinction effect by a varying absorption column
density. The higher $N_H$ of the southern half suppresses more soft X-ray flux
than in the northern half, resulting in the observed colour gradient.

\begin{figure}[t]
    \begin{center}
    \includegraphics[width=0.999\hsize]
    {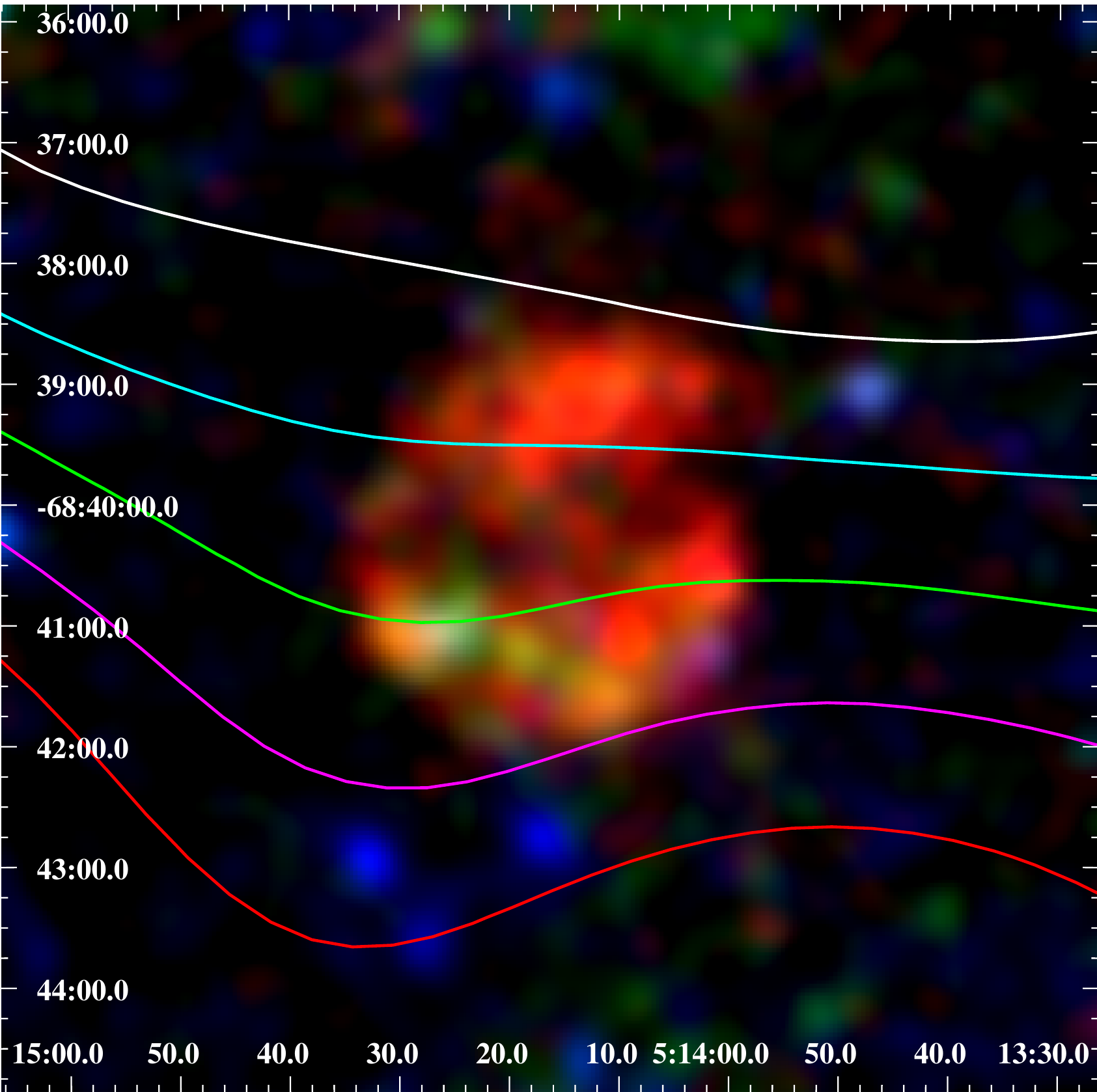}
    \end{center}
\caption{X-ray image of \snrc\ overlaid with \ion{H}{I} column density
contours. Levels shown are 1.5, 1.75, 2.0, 2.25, and 2.5, in units of $10^{21}$
cm$^{-2}$, increasing from north to south.}
    \label{fig_snrc_HI}
\end{figure}

Direct evidence for the north-south density gradient can be found at longer
wavelengths, e.g. in the \ion{H}{I} map of
\citet[][Fig.\,\ref{fig_snrc_HI}]{2003ApJS..148..473K}. The Spitzer MIPS 24
\textmu m image of the  neighbourhood of the remnant
(Fig.\,\ref{fig_rgb_image_snrc}), which traces cold dust, also shows a dustier
environment towards the south. It is reasonable to postulate that the X-ray dark
lane seen across the remnant is also a consequence of foreground absorption,
although the obscuring structure falls below the spatial resolution of the
\ion{H}{I} map.

\paragraph{}
\snrd\ exhibits a more dramatic asymmetry\,: a cool ($\sim 0.1$ keV) X-ray shell
is seen in the SW, correlating with relatively bright H$\alpha$ and
[\ion{S}{ii}] emission. In the NE, we see a ``ridge'' of X-ray emission with
higher temperature ($\sim 0.6$ keV) and little to no co-spatial optical
emission.

\begin{figure*}[t]
    \begin{center}
    \includegraphics[width=\hsize]{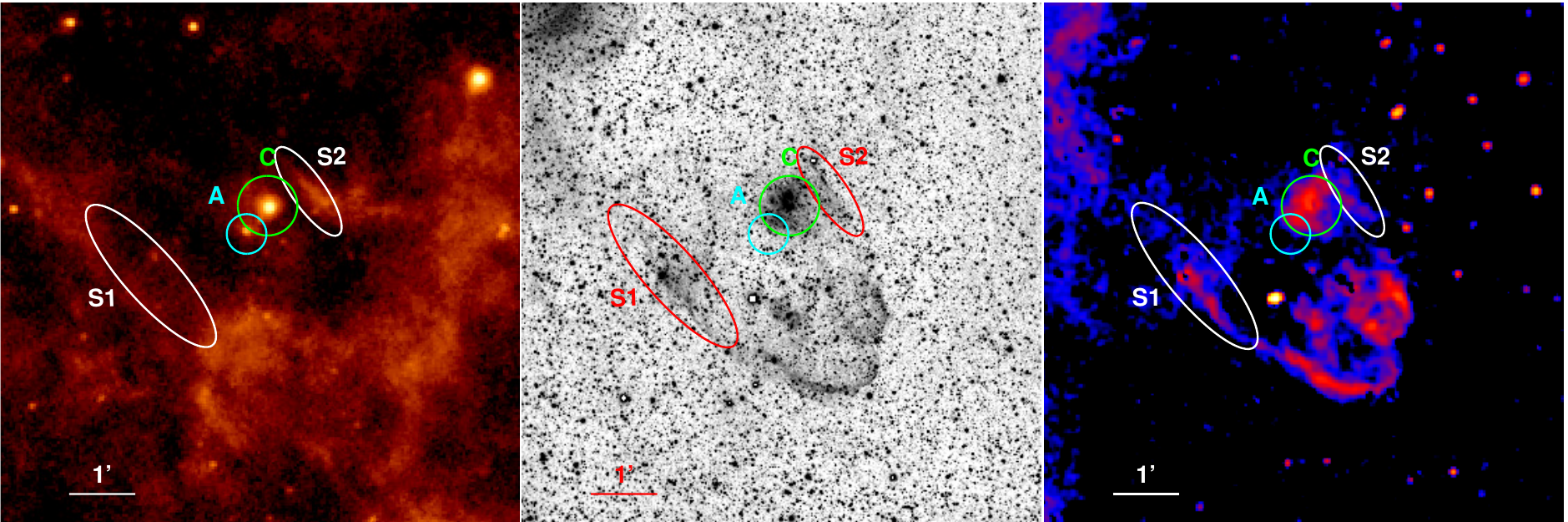}
    \end{center}
\caption{Annotated view of \snrd, as seen at 24 \textmu m (left) and in the
lines
of H$\alpha$ (middle) and [\ion{S}{ii}] (right). The H$\alpha$ image has been
taken with the Blanco \mbox{4-m} telescope and have a pixel size of 0.27\arcsec
$\times$ 0.27\arcsec\ (no continuum has been subtracted). The ellipses (``S1''
and ``S2'') mark faint filaments (see Sect\,\ref{discussion_asymmetric} for
details). The green circle (``C'') shows an unrelated compact \ion{H}{II}
region, and the cyan circle (``A'', left image) marks the background AGN
detected in X-rays and in the mid-IR.}
    \label{fig_snrd_mipdf}
\end{figure*}

To provide a better view of the remnant's emission at various wavelengths, we
show in Fig.\,\ref{fig_snrd_mipdf} an annotated triptych of \snrd\ as seen at 24
\textmu m and in H$\alpha$ and [\ion{S}{ii}] lines. Mid-IR point sources within
the
remnant are the background AGN discussed in Sect\,\ref{results_morphology_snrd}
(cyan circle ``A'' in Fig.\,\ref{fig_snrd_mipdf}) and a $B = 14.75$ mag star
identified as \object{2MASS J05170629-6758401}
\citep{2004AJ....128.1606Z,2006AJ....131.1163S}. The star is likely to simply
lie in projection within \snrd\ and to be unrelated to the remnant. However, the
ionising radiation of 2MASS J05170629-6758401 is responsible for the compact
\ion{H}{II} region around the star that is seen in optical lines (green circle
``C'' in Fig.\,\ref{fig_snrd_mipdf}) and might hide actual SNR emission. In
spite of these interlopers, it is possible to see faint filaments in optical
lines, identified by the ``S1'' and ``S2'' ellipses in
Fig.\,\ref{fig_snrd_mipdf}, which connect the bright optical arc in the SW to
the
X-ray ridge in the NE. 24 \textmu m emission is also seen in these filaments,
possibly originating from compressed, heated dust in the pre-shock region.
Since the temperature is so low (almost too cool to emit X-rays), line emission
by [\ion{O}{iv}] (at 25.9 \textmu m) is also likely contributing to the MIPS
data.
Only the X-ray ridge remains not enclosed by longer wavelength emission.

\begin{figure}[t]
    \begin{center}
    \includegraphics[width=0.999\hsize]{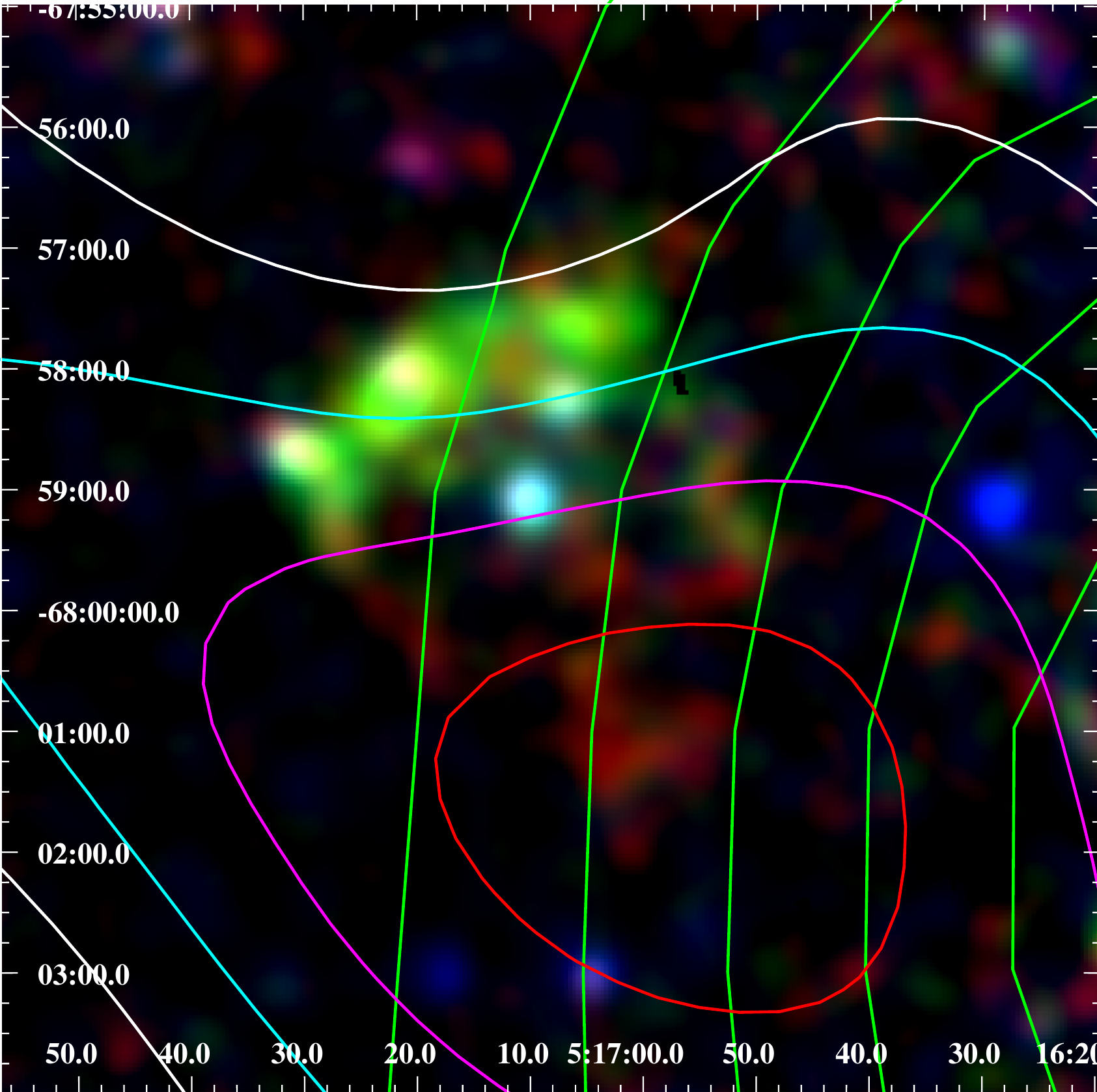}
    \end{center}
\caption{Same as Fig.\,\ref{fig_snrc_HI} for \snrd, with CO contours from the
NANTEN survey added in green. The \ion{H}{I} levels shown are 2.0, 2.25, 2.5,
and 2.75 (in units of $10^{21}$ cm$^{-2}$), in white, cyan, magenta, and red,
respectively. CO contours are from 1$\sigma$ to 5$\sigma$, in steps of 1$\sigma$
($\approx 0.4$ K\,km\,s$^{-1}$), and increase westwards.}
    \label{fig_snrd_HI}
\end{figure}

We again postulate that the asymmetry is essentially governed by the
inhomogeneous ISM. Several tracers indicate a much denser ISM towards the SW of
the remnant, namely\,:
\begin{enumerate}
    \item atomic hydrogen \citep{2003ApJS..148..473K}, see contours on
Fig.\,\ref{fig_snrd_HI}
    \item CO emission; \snrd\ is located at the NE boundary of the giant
molecular cloud (GMC) \object{[FKM2008] LMC N J0516-6807} \citep[][green
contours on Fig.\,\ref{fig_snrd_HI}]{2008ApJS..178...56F}. We stress that the
NANTEN survey only has moderate resolution\footnote{
None of our objects are fully covered by the MAGMA survey of the LMC
\citep{2011ApJS..197...16W}, which has a higher angular resolution ($\sim
45$\arcsec)
}
(beam size of 2.6\arcmin). The green contours on Fig.\,\ref{fig_snrd_HI} are
somewhat misleading, as the brightest part of the GMC (i.e. where most of the
molecular material resides) is really towards the SW, in the same direction as
the elongation of the SNR.
    \item Cold dust; Mid-IR emission (24 \textmu m) outlines the outer part of
the
GMC (can be seen in Fig.\,\ref{fig_snrd_mipdf}).
\end{enumerate}

As \citet{2013arXiv1306.2638L} shows, SNRs interacting with molecular clouds
are the most elliptical (or elongated) remnants. It is easily conceivable that
such remnants will ``forget'' their explosive origins and intrinsic
(a)symmetries
early in their evolution. \snrd\ exemplifies the key role of environment on
shaping the morphology of evolved SNRs.

The evolution of SNRs expanding in a non-uniform ISM has been investigated by
several authors in numerical simulations
\citep[e.g.][]{1996ApJ...471..279D,1999A&A...344..295H,2009MSAIS..13...97O},
pointing to the asymmetries that develop along the density gradient. However,
these studies dealt with much younger SNRs than \snrd. In particular,
\citet{1999A&A...344..295H} showed the strong X-ray surface brightness contrast
that can be produced by density gradients. The brightest emission is expected
from the densest region, as $L_X$ scales with the square of the density. This
effect likely contributes to the slightly brighter emission in the south of
\snrc, but we see the \emph{opposite} trend in the case of \snrd, which shows
brighter emission from the lower density region. This can be interpreted as a
later-time evolution effect\,: The \emph{interaction} of the blast wave with the
much denser ISM in the SW caused the shock to cool down quickly and to become
radiative, leading to a lower level of X-ray emission. The X-rays emitted are
also softer and more easily absorbed (especially since $N_H$ is larger in the
SW), further reducing the observed X-ray flux. This scenario also explains the
stronger optical emission in the SW.

The low-temperature component in the SW is seen essentially as a line complex at
0.5--0.6 keV. There could be charge exchange (CX) emission contribution to that
line complex, instead of a purely thermal emission. CX X-ray emission in SNRs
has been detected in the Cygnus Loop and Puppis A remnants
\citep{2011ApJ...730...24K,2012ApJ...756...49K}. The fraction of CX to the X-ray
emission can be enhanced if the hot gas is interacting with denser, neutral gas,
as we suggest for \snrd. However, the data available allow neither to rule out
the presence, nor to constrain the contribution of CX to the X-ray emission of
the cool SW shell.

The NE part of the blast wave, on the other hand, expands in a more tenuous
environment and is seen as the X-ray ridge. The higher shock temperature (or
expansion velocity) in the lower density region is consistent with the results
of \citet{1996ApJ...471..279D}. As a consequence of the bi-lateral velocity
structure, the apparent centre of the remnant shifts away from the actual
explosion site, towards the lower-density region. The shifts predicted by
\citeauthor{1996ApJ...471..279D} are up to 20\% of the apparent radius. The SN
that created \snrd\ should conversely have exploded south-west of the apparent
geometrical centre, i.e. more embedded within the molecular cloud [FKM2008] LMC
N J0516-6807. This makes the case for a massive stellar progenitor more likely
(see Sect.\,\ref{discussion_typing}).

An alternative explanation might be that the SNR shock broke out into the
lower-density medium in the NE, as suggested for e.g. \object{N11L} and
\object{N86} \citep{1999ApJ...514..798W}. However, the higher temperature of the
plasma in the breakout region and the current lack of detected optical filaments
``streaming'' ahead of the breakout are at odds with this scenario. It is also
possible that \snrd\ is similar to e.g. \object{DEM L238}, \object{DEM L249}, or
\object{MCSNR J0508-6902} \citep[][]{2006ApJ...652.1259B,bozzetto2013}: the
X-ray ``ridge'' is actually at the centre of the SNR shell but only the SW part
of the shell remained detectable. Unless deeper data are available, any scenario
will remain highly speculative and we favour the simpler scenario of
``asymmetric evolution'' described above.

\subsection{Typing the SNRs}
\label{discussion_typing}
\snra\ and \snrb\ are iron-rich, as revealed by the X-ray analysis
(Sect.\,\ref{results_spectroscopy}). Under the assumption that the emission is
composed primarily of ejecta, we derived in
Sect.\,\ref{results_spectroscopy_multicomp} that a large mass of iron ---
between 0.5 \msun\ and 1.2 \msun, depending on the level of hydrogen mixing ---
has been produced in these SNe. However large the uncertainties, these results
clearly favour a type Ia SN origin for \snra\ and \snrb. Indeed, thermonuclear
SNe produce much more iron than CC SNe \citep[e.g.][]{1999ApJS..125..439I}. The
nucleosynthesis yield of core-collapse SNe, on the other hand, peaks for oxygen
and subsequent $\alpha$-group elements \citep[i.e. Ne, Mg, see
e.g.][]{1995ApJS..101..181W}. Therefore, the large Fe/O ratio observed is also
arguing against a CC SN origin. 

\paragraph{}
No ejecta emission is detected from \snrc\ and \snrd, making the study of the
remnants' environment the only path to a tentative typing of their parent SNe.
We attempt to quantify $r=N_{\rm CC}/N_{\rm Ia}$, the ratio of CC SNe to
thermonuclear SNe expected from the observed distribution of stellar ages in the
neighbourhood of the remnants.

Over the visibility time of a remnant --- taking 100 kyr as a very conservative
limit --- the stars in the SFH cell including the remnant will not drift away.
In other words, the distribution of stellar ages observed \emph{now} is the same
as that when the SN exploded. We used the delay time distribution $\Psi _{i}
(\tau)$, the SN rate at time $\tau$ following a star formation event, measured
by \citet{2010MNRAS.407.1314M} in the Magellanic Clouds, with $i=1$, 2, and 3
designating the time intervals they used ($t <$ 35~Myr, 35~Myr $< t <$ 330~Myr,
and 330~Myr $< t < 14$~Gyr, respectively).  From timescale arguments it is
reasonable to assume that $\Psi _1$ will correspond to the CC SN rate, whilst
$\Psi _2$ and $\Psi _3$ will be that of SNe Ia (regardless of their ``prompt''
or ``delayed'' nature). We can integrate the SFR to obtain $M_{i}$, the stellar
mass formed in each time interval. Then, we can compute $r$ as the ratio of the
\emph{rates} of CC and Ia SNe, since the visibility times are the same for both
types, leading to\,:
\begin{equation}
    r = \frac{\Psi_1 M_1}{\Psi_2 M_2 + \Psi_3 M_3}
\end{equation}

The SFHs around \snra\ and \snrb\ (Sect.\,\ref{results_sfh}) lack a significant
peak of recent (less than 40 Myr) star formation activity, which translates
into rather small $r$ values of 1.4 and 0.6, respectively, for these remnants.
A similar value ($r = 1.4$) is obtained for \snrc, as expected given the
similarities of the SFHs of \snra\ and \snrc. Finally, for \snrd, we derived a
much higher ratio of $r = 7.6$.

Intuitively, any value $r > 1$ should favour a CC SN origin (conversely for a
thermonuclear origin). However, an important caveat to interpret these results
is that the rates of \citet{2010MNRAS.407.1314M}, especially $\Psi_2$ and
$\Psi_3$, are quite uncertain, due to the still limited sample of SNRs.
Specifically, $\Psi_2$ has a value that changes by a factor of four depending on
the tracer used to constrain the SNR visibility time. (Here we adopted $\Psi_2 =
0.26$ SNe yr$^{-1}$ ($10^{10}$\msun)$^{-1}$ and $\Psi_3 < 0.0014$ SNe yr$^{-1}$
($10^{10}$\msun)$^{-1}$. Note that because $\Psi_3$ is an upper limit, $r$ is
formally a lower limit.)

To provide a better feeling on what $r$ value to expect in either case, we
derived this ratio for SNRs having a well secured type. We took the sample of
four CC and Ia SNRs of \citet{2009ApJ...700..727B}. For type Ia remnants, we
obtained $r=$ 2.2, 0.9, and 1.0 for DEM~L71, SNR~0509-67.5, and SNR~0519-69.0,
respectively. N103B has $r=3.2$ but is atypical, as it is in a star forming
region. Core-collapse SNRs have $r$ ranging from 2.6 (N49) to more extreme
values (e.g. $r=21.7$ for N158A).

In spite of large uncertainties, the ratio $r=N_{\rm CC}/N_{\rm Ia}$ still
appears as a useful tool to assign a type to SNRs using the observed local SFH.
It provides a quantitative criterion to confirm the intuition that SNR without a
prominent peak of recent star formation are more likely to originate from a type
Ia SN. Values in excess of 2 favour a CC scenario, whilst type Ia SNRs will
exhibit $r \lesssim 1$. $r \gg 2$ or $\ll 1$ allow more secure typing in either
ways. Intermediate value (between 1 and 2) lead to uncertain classification,
though preliminary results (e.g. for \snra) show that a type Ia remnant can have
$r=1.4$.

To conclude, this analysis makes the case for a type Ia origin of \snra\ and
\snrb\ even stronger. \snrd\ is the only source having a prominent peak of
recent star formation. The large $r$ value of 7.6 strongly argues for a CC SN
origin of \snrd. The asymmetric/elongated morphology, likely originating from
the interaction with a molecular cloud, with which the remnant is associated
(Sect.\,\ref{discussion_asymmetric}), is consistent with this type
\citep{2013arXiv1306.2638L}. Finally, we marginally favour a thermonuclear
origin \snrc. We stress, however, that the classification of this remnant is
more uncertain than for the other objects, for which other clues to the SN type
are available.

\subsection{Two iron-rich SNRs}
\label{discussion_comparison}

\snra\ and \snrb\ are remarkable in their morphological and chemical structure.
The centrally located iron-rich plasma betray their type Ia SN origin. They
join the sample of MC SNRs with a faint, soft X-ray shell containing iron-rich
hot gas in collisional equilibrium (see references in theiIntroduction). They
are likely more evolved versions of the very similar remnants DEM~L238,
DEM~L249, and MCSNR J0508$-$6902. The main difference with these is the absence
in our data of a detected X-ray shell, whilst very dim [\ion{S}{ii}] emission
still indicates the locations of the furthest advance of the SN blast wave. This
echoes other cases in the SMC, namely DEM~S128, IKT~5, and IKT~25
\citep{2004A&A...421.1031V}. The faint sulphur emission and lack of soft X-rays
from the shells of \snra\ and \snrb\ indicate that they reached the point when
radiative cooling caused the shells to become either too cool to emit X-rays,
or too faint to be detected in the data available. If the temperature is that
low, we might expect [\ion{O}{iv}] emission at 25.9 \textmu m. The detection of
faint filament in the MIPS image, at the south-eastern rim of the [\ion{S}{ii}]
shell of \snrb, lends support to that scenario.

As discussed in \citet{2006ApJ...652.1259B}, the long ionisation ages of the
iron-rich central plasma is puzzling given the type Ia classification of these
remnants, because it requires higher densities in the centre than expected from
standard type Ia SN models. A possibility is that the SN progenitor was more
massive ($\sim$ 3 to 8 M$_{\sun}$) than average, because the stellar winds of
these ``prompt'' progenitors increase the circumstellar density.

The central location of most of the X-ray emission is reminiscent of the
mixed-morphology SNRs classification
\citep[MMSNRs,][]{1998ApJ...503L.167R,2006ApJ...647..350L}, which has been
applied only to Galactic remnants. The MMSNRs are usually close to, and
interacting with, a molecular cloud environment, showing OH masers. It means
that the MMSNRs very likely have massive progenitors and are not type Ia SNRs,
as we conclude for \snra\ and \snrb.

The number of known iron-rich SNRs with centrally-peaked in X-rays has
significantly increased thanks to the \xmm\ survey of the LMC \citep[][this
work]{bozzetto2013}. We have access to a sample of evolved
remnants which cannot be studied in the Milky Way, because the high column
density towards the Galactic plane, (much) in excess of $10^{22}$ cm$^{-2}$,
readily absorbs most or all the photons below 1 keV. As such, all iron lines are
hidden (there is no Fe K emission at these temperatures). Unfortunately, because
\snra\ and \snrb\ are (relatively) X-ray dim and have only been observed in
short or moderately long exposures, the data currently available  suffer from
low statistics, which is further reduced by the off-axis location of the source
in the survey observations. Deeper observations with the current X-ray
telescopes or future instrumentation \citep[e.g.
Athena+,][]{2013arXiv1306.2335D} are warranted to yield higher-quality images
and spectra. This will enable, e.g., more accurate measurements of the physical
conditions in the plasma, of abundances, or allow to look for spectral
variations across the remnants.

\section{Summary}
\label{summary}
We have presented four new members of the LMC SNR population, discovered in the
survey performed  with \xmm. All four are relatively evolved remnants. Their
thermal X-ray emission is the key signature of their nature, although
observations at longer wavelengths are essential to understand the conditions
resulting in their current appearance. Our analysis concluded that\,:
\begin{enumerate}
\item \snra\ and \snrb\ are iron-rich, centrally-peaked in X-rays SNRs. They
are spectrally very similar to DEM L238, DEM L249, and MCSNR J0508-6902,
although they have no detected shell around the central iron-rich plasma.
However, faint [\ion{S}{ii}] emission surrounding them is likely a relic of the
propagation of the SN blast wave. From X-ray spectral fits, we measure the
presence of (0.5--1) \msun\ and (0.6--1.2) \msun\ of Fe in \snra\ and \snrb,
respectively. The large amounts of iron in the SNR interiors and the lack of
recent star formation in the neighbourhood of the remnants both strongly point
to a type Ia SN origin.

\item \snrc\ is a large diameter (53 pc), roughly spherically symmetric
remnant, whose X-ray spectrum is entirely from the shocked ISM. It displays
optical emission coextensive with the X-rays in the southern half of the
remnant. \snrc\ is the only object of our sample to be detected at radio
frequencies, just above the noise limit, and only from the southern half. This
indicates that radio emission from the other remnants, if present, is too faint
to be detected in currently available data. The north-south asymmetry of the
radio and optical emission is interpreted as a non-uniform ambient medium,
having a density increasing towards the south. The X-ray-derived $N_H$ in the
southern half of the remnant is indeed higher than in the northern one, in
keeping with the $N_H$ gradient seen in the \ion{H}{I} map of the LMC, lending
support to our interpretation.

\item \snrd\ exhibits a very unusual morphology. A very soft X-ray shell in the
south-west is closed by a harder X-ray ``bar'' in the north-east. Optical
emission with high \siiha\ correlates only with the soft X-rays in the
south-west, where a giant molecular cloud is reported. This leads us to the
interpetation that the SN blast wave is propagating into a much denser ISM
towards the south-west. There, the shock quickly slowed and cooled down,
producing the very soft X-rays, whilst being relatively unperturbed towards the
north-east. The association with a molecular cloud and the recent star
formation in the local SFH suggest a massive progenitor for \snrd.

\end{enumerate}

\begin{acknowledgements}
The \xmm\ project is supported by the Bundesministerium f\"ur Wirtschaft und
Technologie\,/\,Deutsches Zentrum f\"ur Luft- und Raumfahrt (BMWi/DLR, FKZ 50 OX
0001) and the Max-Planck Society.
Cerro Tololo Inter-American Observatory (CTIO) is operated by the Association of
Universities for Research in Astronomy Inc. (AURA), under a cooperative
agreement with the National Science Foundation (NSF) as part of the National
Optical Astronomy Observatories (NOAO). We gratefully acknowledge the support of
CTIO and all the assistance which has been provided in upgrading the Curtis
Schmidt telescope.
The MCELS is funded through the support of the Dean B. McLaughlin fund at the
University of Michigan and through NSF grant 9540747.
We used the {\sc karma} software package developed by the ATNF. The Australia
Telescope Compact Array is part of the Australia Telescope which is funded by
the Commonwealth of Australia for operation as a National Facility managed by
CSIRO. 
P.\,M. and P.\,K. acknowledge support from the BMWi/DLR grants FKZ 50 OR 1201
and 50 OR 1209, respectively. M.\,S. acknowledges support by the Deutsche
Forschungsgemeinschaft through the Emmy Noether Research Grant SA 2131/1.
This research has made use of Aladin, SIMBAD and VizieR, operated at the CDS,
Strasbourg, France.
\end{acknowledgements}


\end{document}